\documentclass[aps,prb,twocolumn,amsmath,amssymb,superscriptaddress,showpacs,longbibliography]{revtex4-1}

\usepackage{hyperref}
\pdfoutput=1
\usepackage{graphicx}
\usepackage{epsfig}
\usepackage{epstopdf}

\usepackage{amsmath}
\usepackage{bm}
\usepackage{amssymb}
\usepackage{textcomp}
\usepackage{ulem}

\usepackage{import}
\usepackage{color}
\usepackage[english]{babel}
\usepackage{float}

\begin{document}

\title{All-optical blister test of suspended graphene using micro-Raman
    spectroscopy}

\author{Dominik Metten}
\affiliation{Institut de Physique et Chimie des Mat\'eriaux de Strasbourg and NIE, UMR 7504, Universit\'e de Strasbourg and CNRS, 23 rue du L\oe{}ss, BP43, 67034 Strasbourg Cedex 2, France}

\author{Fran\c{c}ois Federspiel}
\affiliation{Institut de Physique et Chimie des Mat\'eriaux de Strasbourg and NIE, UMR 7504, Universit\'e de Strasbourg and CNRS, 23 rue du L\oe{}ss, BP43, 67034 Strasbourg Cedex 2, France}

\author{Michelangelo Romeo}
\affiliation{Institut de Physique et Chimie des Mat\'eriaux de Strasbourg and NIE, UMR 7504, Universit\'e de Strasbourg and CNRS, 23 rue du L\oe{}ss, BP43, 67034 Strasbourg Cedex 2, France}

\author{St\'ephane Berciaud}
\email{stephane.berciaud@ipcms.unistra.fr}
\affiliation{Institut de Physique et Chimie des Mat\'eriaux de Strasbourg and NIE, UMR 7504, Universit\'e de Strasbourg and CNRS, 23 rue du L\oe{}ss, BP43, 67034 Strasbourg Cedex 2, France}


\begin{abstract}
We report a comprehensive micro-Raman study of a pressurized suspended graphene membrane that hermetically seals a circular pit, etched in a Si/SiO$_2$ substrate. Placing the sample under a uniform pressure load results in bulging of the graphene membrane and subsequent softening of the main Raman features, due to tensile strain. In such a microcavity, the intensity of the Raman features depends very sensitively on the distance between the graphene membrane and the Si substrate, which acts as the bottom mirror of the cavity. Thus, a spatially resolved analysis of the intensity of the G- and 2D-mode features as a function of the pressure load permits a direct reconstruction of the blister profile. An average strain is then deduced at each pressure load, and Gr\"{u}neisen parameters of $1.8\pm0.2$ and $2.4\pm0.2$ are determined for the Raman G and 2D modes, respectively. In addition, the measured blister height is proportional to the cubic root of the pressure load, as predicted theoretically. The validation of this scaling provides a direct and accurate determination the Young's modulus of graphene with a purely optical, hence contactless and minimally invasive, approach. We find a Young's modulus of $\left(1.05\pm 0.10\right) \rm TPa$ for monolayer graphene, in a perfect match with previous nanoindentation measurements. This all-optical methodology opens avenues for pressure sensing using graphene and could readily be adapted to other emerging two-dimensional materials and nanoresonators.


\end{abstract}

\maketitle
\newpage

\section{Introduction}

Two-dimensional crystals \cite{Novoselov2005}, being just one or a few atoms thick and having lateral dimensions ranging from micrometers up to macroscopic scales, are a new class of solid state membranes. Among these systems, graphene has attracted considerable interest, due to its unique electronic band structure \cite{Castroneto2009} as well as its outstanding materials properties. In particular, graphene is endowed with exceptional mechanical properties, such as a large Young's modulus and intrinsic strength \cite{Lee2008}, ultrastrong adhesion \cite{Koenig2011}, and impermeability to standard gases \cite{Bunch2008}. Owing to the great electrical controllability of graphene \cite{Novoselov2004}, suspended graphene membranes can conveniently be integrated into nanoelectromechanical resonators \cite{Bunch2007,Chen2009}. In addition, graphene interacts strongly with optical radiation \cite{Koppens2011}. However, being atomically thin, a single layer of graphene is quasitransparent over the infrared and visible ranges \cite{Nair2008,Mak2008}. These features allow the optical readout of mechanical resonances \cite{Bunch2007,Bunch2008,Castellanos-Gomez2013,Reserbat2012} and open perspectives for optomechanical studies \cite{Barton2012}. 

It was also recently demonstrated that the impermeability and ultrastrong adhesion of graphene make it possible to form blisters (or balloons), by applying a pressure difference between both sides of a suspended graphene membrane \cite{Bunch2008}. Such systems are highly promising for molecular sieving applications \cite{Koenig2012}. In practice, bulging of the atomically thin membrane can be quantitatively investigated by using atomic force microscopy (AFM) \cite{Bunch2008,Koenig2011} or nanoindentation \cite{Lee2008}, in what is known as a blister (or bulge) test \cite{Komaragiri2005}. In addition, the resulting strain field in the bulged graphene membrane may be probed optically, through frequency shifts of the main Raman scattering features \cite{Georgiou2011,Zabel2011,Lee2012,Kitt2013}. A quantitative analysis requires, however, knowledge of the Gr\"{u}neisen parameters, whose determination is challenging in pressurized suspended graphene membranes. 

Here, we show that micro-Raman scattering alone not only permits one to investigate strain-induced phonon softening in pressurized graphene membranes, but also readily provides the blister topography, resulting in a comprehensive, all-optical blister test. The height profile of a pressurized graphene blister is determined from the analysis of the integrated intensity of the main (G and 2D) Raman scattering features of graphene. This allows a direct determination of the tensile strain. The softening of the Raman features is then examined under known tensile strain, as a function of the pressure load. The Gr\"{u}neisen parameters for the G- and 2D-mode features and, importantly, the Young's modulus of graphene are then obtained only by optical means.
This approach is contactless and can thus be applied in a large variety of experimental conditions. It could serve as a guide for further optical and optomechanical studies on graphene and related systems.

\section{Experimental methods}

\begin{figure*}[!th]
\begin{center}
\includegraphics[scale=1.0]{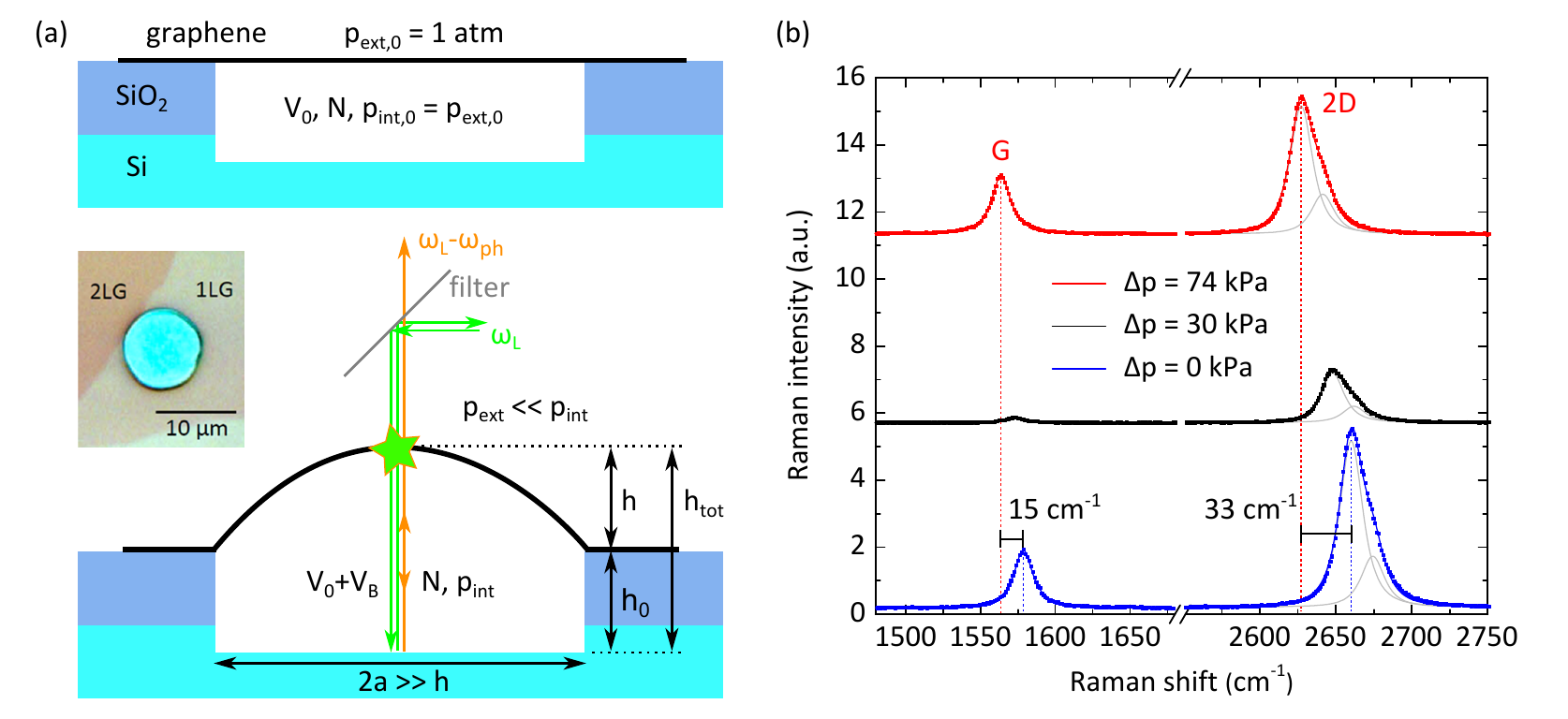}
\caption{Formation of a pressurized graphene blister. (a) Sketch of a suspended graphene membrane at pressure equilibrium (upper part) and under a uniform pressure load (lower part).  An optical image of a suspended graphene monolayer sealing a cylindrical pit with a radius $a\approx 4~\rm \mu m$ is shown in the center left part of (a). (b) Micro-Raman spectra recorded at the center of the graphene membrane at different values of $\Delta p=p_{\rm int}-p_{\rm ext}$.}
\label{fig01}
\end{center}
\end{figure*}

For an accurate blister test, high quality, impermeable and defect-free graphene is mandatory. We therefore prepare our graphene samples by mechanical exfoliation of natural graphite. This material is known to be well suited to the investigation of the intrinsic properties of graphene. Graphene layers are deposited over circular pits that have been patterned in a Si/SiO$_2$ substrate by optical lithography, followed by reactive ion etching and careful drying, in order to eliminate liquid residues inside the pits~\cite{Berciaud2009}. No silicon oxide is left within the pits. The mechanically exfoliated graphene layers are tightly clamped around the border of the pit by van der Waals forces, resulting in a hermetically sealed drum, in which a constant number of air molecules is trapped. The typical pit radius is $a\approx4~\mu$m and the pit depth $h_0$ is measured with a profilometer. In the following, we present results obtained for a sample with $h_0=(395\pm10)~\rm nm$. Similar results are obtained on two other samples with different pit depths (see Supplemental Material~\cite{SMnote}).

As a characterization tool, we make use of micro-Raman scattering spectroscopy \cite{Malard2009,Ferrari2013}, which is highly sensitive to the number of layers, disorder \cite{Cancado2011,Eckmann2012}, doping, and, importantly, to strain \cite{Mohiuddin2009,Huang2009,Metzger2009,Ding2010,Zabel2011,Lee2012,Kitt2013,Lee2012a,Huang2010,Yoon2011,Frank2011}. Since suspended graphene is immune to substrate-induced doping \cite{Berciaud2009, Berciaud2013} and minimally sensitive to atmospheric doping~\cite{Ryu2010}, these samples allow the direct investigation of strain-induced changes in the Raman spectrum of graphene, without spurious contributions from a residual charge-carrier density \cite{Metten2013}. Here, micro-Raman measurements are performed in a backscattering geometry, with a homebuilt setup, by using a $20\times$ objective $(\rm N.A.=0.45)$ and a 532-nm laser beam focused onto an approximately $1.2~$-$\rm \mu m$ (full width at half maximum) spot. The objective is mounted onto a piezoelectric stage allowing spatially resolved Raman studies. The collected Raman scattered light is dispersed onto a charged-coupled device array by a single-pass optical spectrometer, with a spectral resolution better than $2~\rm cm^{-1}$. The laser beam is linearly polarized and the laser power is maintained at 0.7\,mW, in order to avoid laser-induced local heating and subsequent thermally induced spectral shifts or line shape changes of the Raman features \cite{Calizo2007}. Suspended graphene monolayers are unambiguously identified from the characteristic line shape of their Raman 2D-mode feature \cite{Berciaud2013}, and their undoped character is systematically confirmed from a detailed spatially resolved Raman study \cite{Berciaud2009}. The relative integrated intensity of the defect-related D mode to that of the G mode is less than $1~\%$ on the samples investigated here. In order to form graphene blisters, the samples are held in a vacuum chamber equipped with a quartz window for optical access. The external pressure  $p_{\rm ext}$ is smoothly varied from approximately $10^{-2} ~\rm Pa$ to atmospheric pressure $(100\pm2~\rm) kPa$. Considering the bulging of the graphene membrane, we estimate, by using the ideal gas law, that the corresponding pressure load $\Delta p = p_{\rm int}-p_{\rm ext}$, i.e., the difference between pressures inside and outside the blister, varies between $(0~\rm \pm2)$ and $(74~ \pm5)~\rm~kPa$, respectively. More details on the determination of $\Delta  p$ are given in Supplemental Material~\cite{SMnote}. Importantly, the highest $\Delta p$ achieved here is more than one order of magnitude below the threshold, at which delamination occurs \cite{Koenig2011}. Consequently, we will consider a constant blister radius throughout this article.

Figure \ref{fig01} shows a schematic of our experimental approach. The volume of the cylindrical pit is $V_0$, $V_{\rm B}$ denotes the volume of the blister, and $N$ is the number of trapped air molecules. We note  $h(r)\ll a$, the vertical displacement of the graphene layer, and $h_{\rm tot}(r)$, the total distance between graphene and the underlying Si substrate, at a distance $r$ from the center of the blister. The maximum deflection $h(0)$ is denoted $h_{\rm max}$. Considering our sample geometry, we estimate an upper bound for the maximum angle between the substrate and the bulged graphene of approximately $0.1 ~\rm rad$. Therefore, we assume that the laser beam always impinges on the graphene membrane at quasinormal incidence.

Over long timescales, on the order of several hours, graphene blisters tend to deflate, essentially due to slow diffusion of air molecules through the Si/SiO$_2$ substrate \cite{Bunch2008, Koenig2011}.
In order to verify whether the leak rate has to be considered, Raman measurements are performed on suspended graphene membranes at $p_{\rm ext}=p_{\rm int}=100~\rm kPa$ before pumping out the vacuum chamber and again at $p_{\rm ext}=100~\rm kPa$,  after a series of measurements as a function of $p_{\rm ext}$, starting from $p_{\rm ext}\approx 10^{-2} ~\rm Pa$. No significant changes of the Raman frequencies or of the integrated intensity of the Raman features are observed, which demonstrates that the leak rate of our pressurized membrane could be neglected over the duration of a measurement run. Consequently, a constant $N$ is assumed in the analysis described below. Data performed on longer time scales, revealing for a finite leak rate, evidenced by the development of a concave blister profile at the end of a measurement cycle, are shown in Supplemental Material~\cite{SMnote}.

\section{Strain-induced phonon softening}

Raman spectra recorded at the center of the membrane for $\Delta p=0 ~\rm kPa$ and $\Delta p=74 ~\rm kPa$, are shown in Figure \ref{fig01}(b). At pressure equilibrium, the Raman G-mode feature (fit to a single Lorentzian) is centered at $\omega_{\rm G}=1578.8~\rm cm^{-1}$, with a full width at half maximum (FWHM) of $\Gamma_{\rm G}^{}=(15\pm0.5)~\rm cm^{-1}$, characteristic of an undoped sample \cite{Berciaud2009}. Its integrated intensity is denoted $I_{\rm G}$. We note that $\Gamma_{\rm G}$ remains at $(15\pm0.5)~\rm cm^{-1}$ over the suspended membrane at each value of $\Delta p$. This value confirms that doping from the surrounding air molecules can be neglected and that suspended graphene membranes allow investigations of strain without parasitic effects from unintentional doping. 
The 2D-mode feature shows an asymmetric line shape, as typically observed on suspended graphene, and is fit to a modified double Lorentzian profile, as in ref. \cite{Berciaud2013}. The lower energy feature has much higher integrated intensity and its peak frequency coincides with the peak frequency of the 2D-mode feature. The spectral shift between the low- and the high-energy features (approximately $15~\rm cm^{-1}$), as well as their integrated intensity ratio (approximately 3) is also constant over the suspended part, irrespective of $\Delta p$. Hence, we use the position of the low-energy 2D-mode subfeature as the peak frequency, denoted $\omega_{\rm 2D}$, and the sum of the integrated intensities of both subfeatures is referred to as $I_{\rm 2D}$. The fact that values of $\omega_{\rm G}^{}=1578.8$\,cm$^{-1}$ and $\omega_{\rm 2D}^{}=2660.0$\,cm$^{-1}$ are slightly lower than expected for pristine graphene is attributed to an initial built-in strain of less than $0.1\%$, in accordance with our previous studies \cite{Metten2013}. Very similar results to those described below are also obtained on suspended samples, on which no significant built-in strain is observed (see Supplemental Material~\cite{SMnote}). This similarity suggests that prestrain has no major effect on bulging under uniform pressure load, which is consistent with the negligible bending rigidity of graphene \cite{Lee2008,Komaragiri2005}.

\begin{figure*}[!htb]
\begin{center}
\includegraphics[scale=1.0]{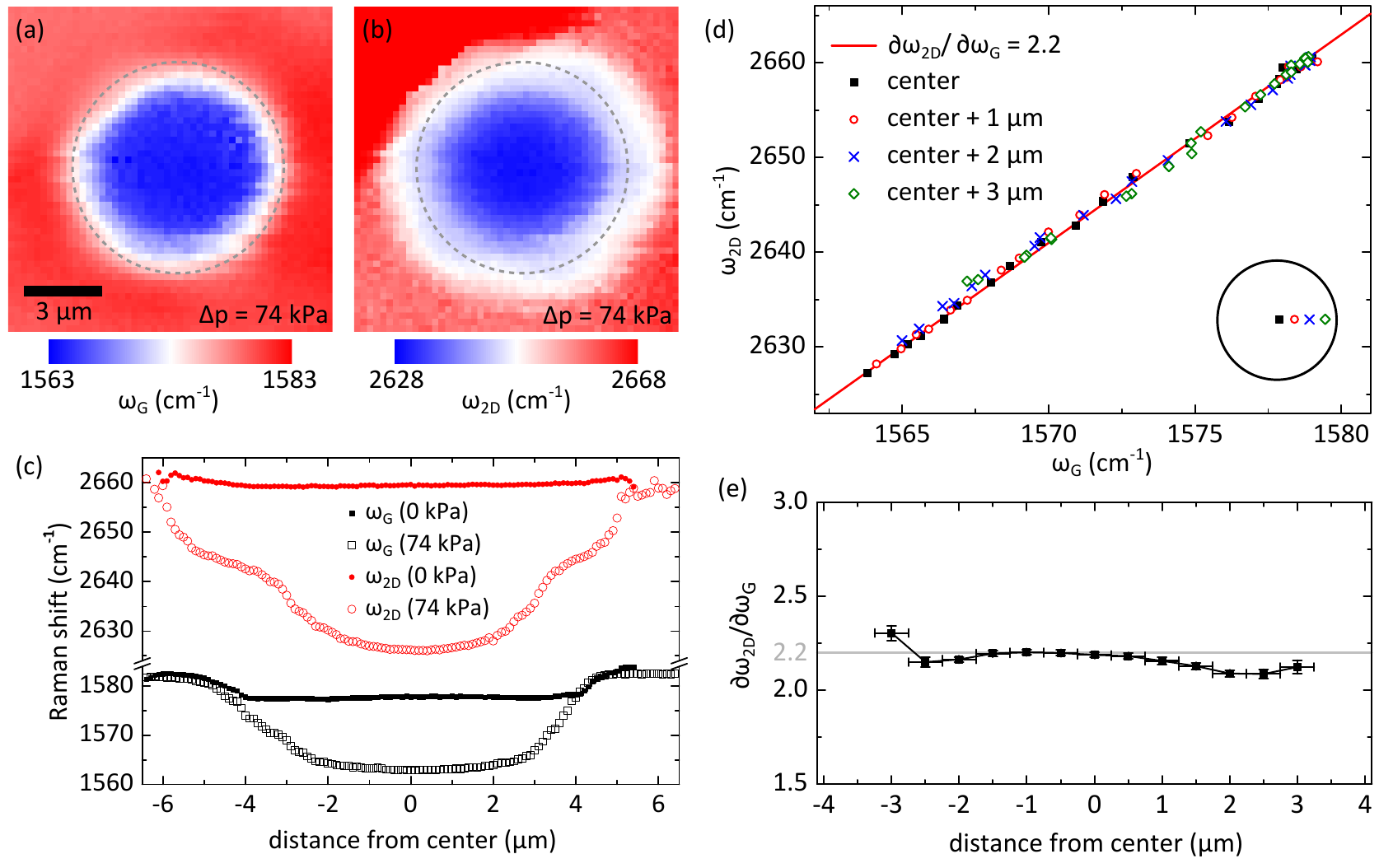}
\caption{Strain-induced phonon softening. (a),(b) Spatially resolved Raman maps of the G- and 2D-mode frequencies recorded on the sample shown in Figure \ref{fig01}(a), under a uniform pressure load of $\Delta p= 74~\rm kPa$. The step size is 250\,nm. The upper left part of the sample contains a supported bilayer region, where, as expected, $\omega_{\rm 2D}^{}$ upshifts significantly. The border of the pit is represented by gray dashed circles. (c) High-resolution radial line scans of the frequencies of the G- (black squares) and 2D-mode (red circles) features, recorded across the pit at $\Delta p= 0~\rm kPa$ (filled symbols) and $\Delta p= 74~\rm kPa$  (open symbols), with a step size of 100\,nm. (d) Correlation between $\omega_{\rm G}$ and $\omega_{\rm 2D}$ plotted for each pressure difference at four different values of $r$, ranging from $r=0~\rm \mu m$ to $r=3~\rm \mu m$. The solid line is a linear fit with a slope $\partial \omega_{\rm 2D}^{}/\partial \omega_{\rm G}=2.2$. (e) $\partial \omega_{\rm 2D}^{}/\partial \omega_{\rm G}$ as a function of $r$, the distance from the blister center.}
\label{fig02}
\end{center}
\end{figure*}

When placing the sample under high vacuum [see red curve in Figure \ref{fig01}(b)], both the G- and 2D-mode features soften (by 15\,cm$^{-1}$ and 33\,cm$^{-1}$ at the center of the membrane, respectively) but retain their peak shapes and show comparable values of $I_{\rm G}$ and $I_{\rm 2D}$. A spectrum taken at an intermediate $\Delta p=30 ~\rm kPa$ is also shown. Interestingly, it reveals a striking decrease of $I_{\rm G}$, by one order of magnitude, and of $I_{\rm 2D}$ by a factor of only approximately 4, compared to the measurement at $\Delta p=0 ~\rm kPa$. These variations are key in our analysis and are discussed later in the manuscript. We first concentrate on the Raman shifts and their dependence on $\Delta p$ and on $r$.

Figure \ref{fig02} displays two-dimensional maps, of $\omega_{\rm G}$ (a) and $\omega_{\rm 2D}$ (b), recorded at $\Delta p= 74 ~\rm kPa$ on the sample shown in Figure \ref{fig01}(a). The pressurized suspended region exhibits centrosymmetric distributions of $\omega_{\rm G}$ and  $\omega_{\rm 2D}$ with minimum values much smaller than on the supported region. Indeed, a few microns away from the pit, the pressure-induced strain is relaxed, and homogeneous distributions of $\omega_{\rm G}=(1581\pm1)$\,cm$^{-1}$, $\omega_{\rm 2D}=(2661\pm2)$\,cm$^{-1}$ and of $\Gamma_{\rm G}=9.5\pm 0.5 ~\rm cm^{-1}$ are observed on supported graphene. The latter value suggests that this region is slightly doped, by approximately $2\times10^{12}~\rm cm^{-2}$, while the values of $\omega_{\rm G}$ and  $\omega_{\rm 2D}$ are consistent with a built-in tensile strain comparable to the one observed on the suspended region at $\Delta p=0~\rm kPa$ \cite{Berciaud2009,Lee2012a,Metten2013}. 

In Figure \ref{fig02}(c), we further compare the G- and 2D-mode frequencies at $\Delta p= 0 ~\rm kPa$ and $\Delta p= 74 ~\rm kPa$ along a radial line scan across the pit. For both data sets, the measured G-mode frequencies converge very near the border of the pit (at approximately $4~\rm \mu m$ from the center), whereas for $\omega_{\rm 2D}$ the convergence is observed at $\approx 5.5~\rm \mu m$ from the center of the pit. We attribute this difference to the subtle interplay between the evolution of $\omega_{\rm G}$ and $\omega_{\rm 2D}$, due to strain relaxation at the edges of the pit, and the presence of residual doping on the supported part \cite{Lee2012a}. These effects will be discussed in detail elsewhere, since we are interested in studying pure strain on suspended graphene.

To further unveil phonon softening induced by tensile strain, we now investigate the correlation between $\omega_{\rm 2D}$ and $\omega_{\rm G}$ as a function of $\Delta p$ and the position on the graphene membrane. For this purpose, we record Raman line scans with a step size of 500\,nm for 19 different values of $\Delta p$ ranging from 74 down to $0~\rm kPa$. In Figure \ref{fig02}(d) we show the correlation between $\omega_{\rm 2D}$ and $\omega_{\rm G}$ recorded with varying $\Delta p$, at the center of the pressurized membrane, and at $r= 1$, 2 and 3\,$\mu \rm m$ from the center. At $\Delta p =74~\rm kPa$, $\omega_{\rm G}$ $\left(\omega_{\rm 2D}\right)$ shifts down to 1563.8\,cm$^{-1}$ $(2627.2~\rm cm^{-1})$ at the center, whereas $\omega_{\rm G}$ $\left(\omega_{\rm 2D}\right)$ is 1567.7\,cm$^{-1}$ $(2636.0~\rm cm^{-1})$ at 3\,$\mu$m away from the center. As shown in Figure \ref{fig02}(e), when varying $\Delta p$, the correlation between $\omega_{\rm 2D}$ and $\omega_{\rm G}$ is linear, with a slope of $\frac{\partial\omega_{\rm 2D}}{\partial\omega_{\rm G}}=2.2\pm0.1$, irrespective of the position on the suspended graphene blister, within a distance of $3\rm \mu m$ from the center. 

Given the membrane geometry, the pressure-induced stress and resulting tensile strain are essentially biaxial in the pressurized blister~\cite{Zabel2011}. Still, there may be a dominant radial, hence uniaxial, contribution when approaching the edges of the pit \cite{Lee2012}. In our measurements, we observe a splitting of the G-mode feature below 500\,nm from the border, which may arise from uniaxial strain \cite{Huang2009,Mohiuddin2009}. However, the resulting G-mode line shape is independent on the polarization of the incoming and scattered photons (see Supplemental Material~\cite{SMnote}). The apparent bimodal G-mode feature is thus attributed to a superposition of the Raman responses of the supported and suspended regions, due to the finite size of the laser spot, as it has been observed by Lee, Yoon and Cheong \cite{Lee2012}. This result suggests that contributions from uniaxial strain cannot be unambiguously resolved in the present study. Nevertheless, uniaxial or quasiuniaxial strain presumably results in the smaller phonon softening that is observed when approaching the edges of the pressurized membrane, compared to the larger downshifts measured near the center, which arise from biaxial strain. We believe that the levels of strain achieved here are presumably too small to result in a sizable splitting of the Raman features near the edges of the graphene blister.

\section{Reconstruction of the blister topography}

\begin{figure}[!htb]
\begin{center}
\includegraphics[scale=1]{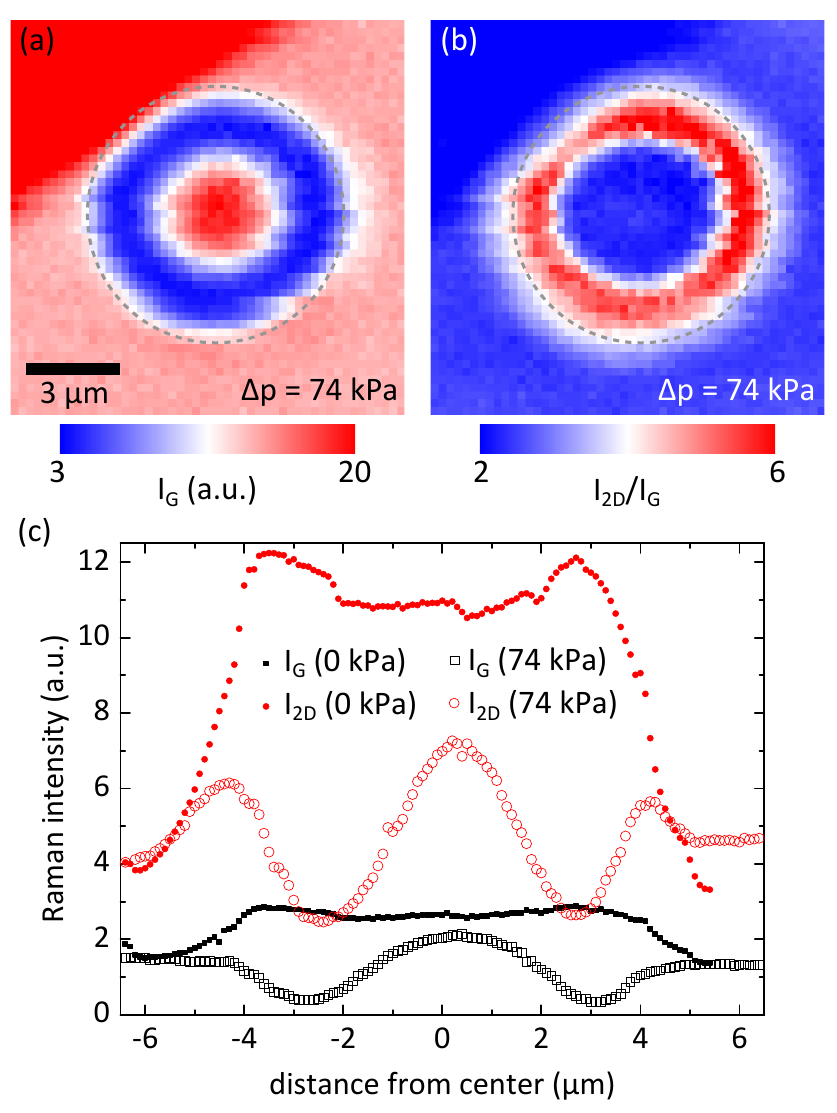}
\caption{Influence of the pressure load on the Raman scattering intensity. Maps of (a) the integrated intensity of the G-mode feature $I_{\rm G}$ and (b) of $I_{\rm 2D}/I_{\rm G}$, the ratio of the integrated intensities of the 2D- and G-mode features recorded on the sample shown in Figure \ref{fig01}(a), under a uniform pressure load of $\Delta p= 74~\rm kPa$. The step size is 250\,nm. The border of the pit is represented by gray dashed circles. (c) High-resolution radial line scans of $I_{\rm G}$ (black squares) and $I_{\rm 2D}$ (red circles), recorded across the pit at $\Delta p= 0~\rm kPa$ (filled symbols) and $\Delta p= 74~\rm kPa$  (open symbols), with a step size of 100\,nm.}
\label{fig03}
\end{center}
\end{figure}

We now address the strong variations of the Raman scattering intensity observed when varying $\Delta p$. Since the Si surface at the bottom of the pit acts as a semireflecting mirror for visible photons, we expect the intensities of the Raman G- and 2D-mode features to depend sensitively on the height of the graphene blister, due to interference effects \cite{Blake2007,Yoon2009,Reserbat-Plantey2013}. Indeed, interference rings appear clearly on the Raman maps of $I_{\rm G}$ and $I_{\rm 2D}/I_{\rm G}$ recorded at $\Delta p = 74 ~\rm kPa$ (see Figure \ref{fig03}(a) and (b). This result demonstrates that $I_{\rm G}$ and $I_{\rm 2D}$ vary significantly over the pressurized membrane and not in the same manner. Conversely, as shown in the line scans of the Raman scattering intensities [see Figure \ref{fig03}(c)], $I_{\rm G}$ and $I_{\rm 2D}$ are nearly constant over the suspended area at $\Delta p =0~\rm kPa$, which is consistent with a nearly flat suspended membrane at pressure equilibrium.

\begin{figure}[!htb]
\begin{center}
\includegraphics[scale=1]{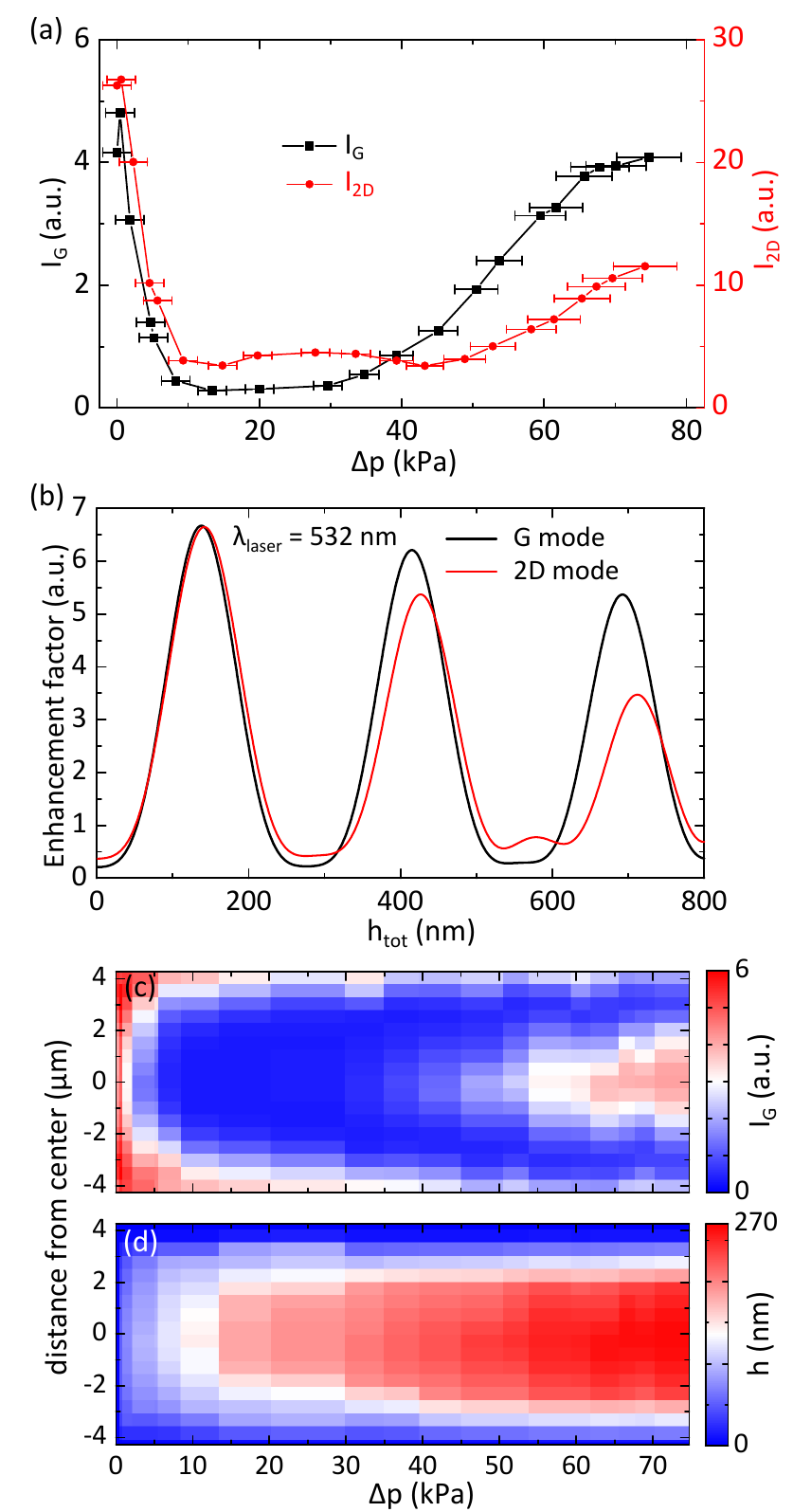}
\caption{Determination of the blister height from the Raman scattering intensity. (a) Evolution of the integrated intensities of the G- and 2D-mode features measured at the center of the sample shown in Figure \ref{fig01}(a), as a function of the pressure load $\Delta p$. (b) Calculated Raman enhancement factors. (c) Raman G-mode intensity $I_{\rm G}$ and (d) blister height $h\left(r\right)$, deduced from the data in (c), as a function of the distance from the blister center $r$ and the pressure load $\Delta p$.}
\label{fig04}
\end{center}
\end{figure}

The evolution of $I_{\rm G}$ and $I_{\rm 2D}$ as a function of $\Delta p$ at $r=0~\rm \mu m$ is represented in Figure \ref{fig04}(a). The ratio between the maximal and minimal value of $I_{\rm G}$ $\left( I_{\rm 2D}\right)$ reaches approximately 13 (approximately 6), and these two quantities are not proportional to each other. The Raman enhancement factor in the graphene-air-silicon layered system can be calculated with a simple analytical model, using the tabulated dielectric constants of Si and bulk graphite, as introduced by Yoon \textit{et al.} \cite{Yoon2009}. Since we use a relatively low numerical aperture objective ($\rm N.A.=0.45$), we assume that the normal incidence approximation is valid in the vicinity of the graphene blister. The key point of this model is that the measured Raman scattering intensity depends not only on the total intensity of the laser beam (at wavelength $\lambda_{\rm laser}$) at the location of the graphene membrane, but also on the total intensity of the backscattered Raman G- and 2D-mode photons at wavelengths $\lambda_{\rm 2D}>\lambda_{\rm G}>\lambda_{\rm laser}$. Both quantities are strongly dependent on $h_{\rm tot}$ (see Figure \ref{fig01}). Consequently $I_{\rm G}$ and $I_{\rm 2D}$ are expected to exhibit distinct evolutions as a function of $h_{\rm tot}$. Thus, from the measured Raman intensities, it is possible to deduce $h_{\rm tot}(r)$ and, finally, the blister height $h\left(r\right)=h_{\rm tot}\left(r\right)-h_0$.

The Raman enhancement factors for the G and 2D Raman modes, computed by using the analytical model of Yoon \textit{et al.} are shown in Figure \ref{fig04}(b) as a function of $h_{\rm tot}$, for $\lambda_{\rm laser}=532$\,nm. We note that, although the values of $h_{\rm tot}$ corresponding to maxima and minima of the enhancement factors are essentially determined by the wavelengths of the laser and Raman scattered photons, the contrast between the maximal and minimal enhancement factors depends sensitively on materials parameters, such as the wavelength-dependent dielectric constants of Si and graphene. This contrast may also be affected by experimental factors, such as local corrugation on the Si surface, as well as slight deviations from the normal incidence approximation, arising from the numerical aperture of the microscope objective or occurring near the edges of the pressurized membrane. Consequently, the calculated enhancement factors are renormalized with respect to the experimentally measured  maxima and minima of $I_{\rm G}$ and $I_{\rm 2D}$. In practice, this renormalization has a minor impact of the determination of $h_{\rm tot}(r)$.

Let us emphasize that, in principle, a simple measurement of the backreflected laser intensity could be employed to deduce $h_{\rm tot}(r)$ \cite{Blake2007,Reserbat2012}. However, due to the quasi transparency of single-layer graphene, the maximum contrast expected in a reflectivity measurement is at most on the order of approximately $15\%$ for a graphene monolayer \cite{Blake2007} , while we obtain a contrast of more than one order of magnitude on $I_{\rm G}$. In addition, Raman measurements also provide quantitative information on the strain field in the graphene blister, as discussed above.

We now compare the data in Figure \ref{fig04}(a) and Figure \ref{fig04}(b). The experimental evolution of $I_{\rm G}$ and $I_{\rm 2D}$ as a function of $\Delta p$ [Figure \ref{fig04}(a)] qualitatively resembles the calculated Raman enhancement factors [Figure \ref{fig04}(b)]. In particular, at $\Delta p=0~\rm kPa$, $h_0=(395 \pm 10)~\rm  nm$,  $I_{\rm G}$ and $I_{\rm 2D}$ are close to their maximum values, which are reached at a finite $\Delta p\approx 1 ~\rm kPa$. This evolution is very consistent with the calculated enhancement factors, which predict maxima at $h_{\rm tot}= 416~\rm nm$ $\left(h_{\rm tot}= 426~\rm nm\right)$ for $I_{\rm G}$ $\left(I_{\rm 2D}\right)$. Similarly, $I_{\rm G}$ and $I_{\rm 2D}$ reach local minima at $\Delta p\approx 14 ~\rm kPa$, corresponding to $h_{\rm tot}\approx 550~\rm nm$ and rise again towards another local maximum at higher $\Delta p$, which would correspond to $h_{\rm tot}= 692~\rm nm$ $\left(h_{\rm tot}= 712~\rm nm\right)$ for $I_{\rm G}$ $\left(I_{\rm 2D}\right)$. This result readily allows us to estimate that the maximum height $h_{\rm max}=h_{\rm tot}(0)-h_0$ of the graphene blister, attained at $\Delta p=74~\rm kPa$, is close to $270~\rm nm$. Interestingly, the evolution of $I_{\rm 2D}$ vs. $\Delta p$ also reveals a slight bump in the range $15~\rm kPa - 45~\rm kPa$, with a secondary maximum around $\Delta p\approx 30~\rm kPa$. This feature also appears clearly in the theoretical calculation of the enhancement factor of the 2D mode near $h_{\rm tot}\approx 580~\rm nm$ (i.e., $h\approx 185~ \rm nm$). This secondary maximum arises from the fact that the Raman enhancement factor is the product of an excitation term, with a quasiperiod of half the laser wavelength and a scattering term, with a larger quasiperiod of half the wavelength of the Raman scattered photons \cite{Yoon2009}. For Raman features at sufficiently large shifts (such as the 2D-mode feature), this \textit{beating} produces secondary maxima in the Raman enhancement factor. Conversely, a significant secondary maximum is neither expected nor observed for $I_{\rm G}$ in the height range investigated here. This observation further validates our experimental approach for the determination of the blister profile.

\begin{figure}[!t]
\begin{center}
\includegraphics[scale=1]{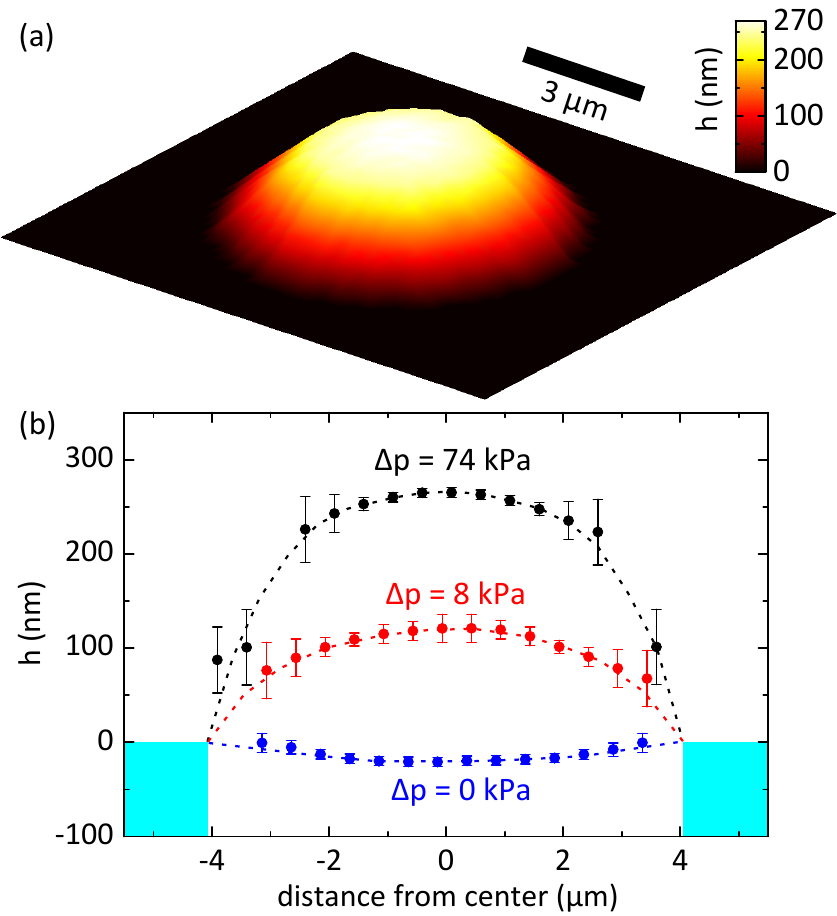}
\caption{Reconstruction of the blister topography. (a) Reconstructed three-dimensional image of the pressurized blister topography at $\Delta p= 74~ \rm kPa$. (b) Blister height profile recorded at various values of $\Delta p$. The error bars in Figure \ref{fig05}(b) take into account the fact that it is not possible to give an accurate value of the height when the Raman intensity is approaching a local minimum [see Figure \ref{fig04}(b)]. The dashed lines are guides to the eye.}
\label{fig05}
\end{center}
\end{figure}

As an example, contour plots of $I_{\rm G}$ and of the corresponding blister height are presented as a function of $\Delta p$ and $r$ in Figs. \ref{fig04}(c) and \ref{fig04}(d). Similar data for $I_{\rm 2D}$ and for another sample are shown in Supplemental Material~\cite{SMnote}. We find that the heights deduced from $I_{\rm G}$ and $I_{\rm 2D}$, respectively, are very similar (see also Figure \ref{fig07}). We are now able to investigate the blister topography in more details. In Figure \ref{fig05}(a), we show a three-dimensional image of the pressurized blister, reconstructed from the Raman map of $I_{\rm G}$ shown in Figure \ref{fig03}(a), by using the approach described above. Cross sections at different values of $\Delta p$ are shown in Figure \ref{fig05}(b). When approaching the border of the circular pit, the measured Raman intensities may be affected by contributions from the neighboring supported graphene. Therefore, the blister profile was linearly interpolated between $r=3~\rm \mu m$ and $r=4.1~\rm \mu m$, where $h_{\rm tot}=h_0$.

\section{Determination of the Gr\"{u}neisen parameters}

Having determined the blister topography, we can now estimate an average tensile strain induced by the uniform pressure load $\epsilon_{\rm p}=L/2a-1$, where $L$ is the length of the cross section of the pressurized graphene blister [see Figure \ref{fig05}(b)]. We find that $\epsilon_{\rm p}$ reaches values of up to $(0.33\pm0.07)~\%$.

\begin{figure}[!htb]
\begin{center}
\includegraphics[scale=1.0]{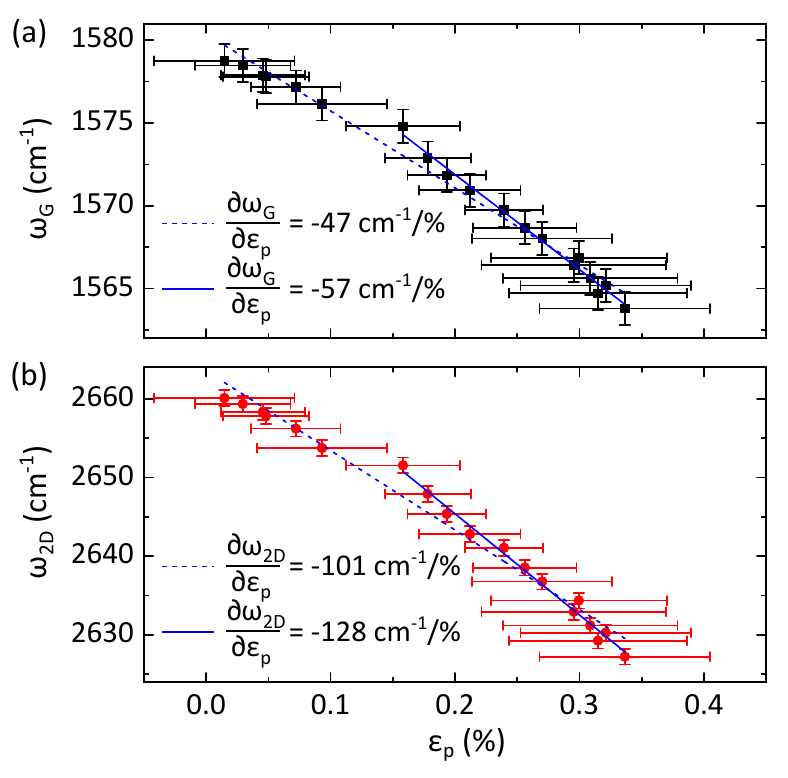}
\caption{Determination of the Gr\"{u}neisen parameters. Evolution of the G-mode (a) and 2D-mode (b) frequencies measured at the center of the sample shown in Figure \ref{fig01}(a), as a function of the tensile strain $\epsilon_{\rm p}$ induced by the uniform pressure load. The straight lines are linear fits.}
\label{fig06}
\end{center}
\end{figure}

We can now correlate $\epsilon_{\rm p}$ to the Raman frequencies $\omega_{\rm G}$ and $\omega_{\rm 2D}$ measured at the center of the blister, as it is shown in Figure \ref{fig06}. Over the range $\epsilon_{\rm p}=0~\%-0.33~\%$, we observe roughly linear scalings with slopes $\partial\omega_{\rm G}/\partial\epsilon_{\rm p}=(-47\pm5)$\,cm$^{-1}$/\% strain and $\partial\omega_{\rm 2D}/\partial\epsilon_{\rm p}=(-101\pm10)$\,cm$^{-1}$/\% strain, respectively. 
Nevertheless, in the limit of small deflections, a precise determination of $\epsilon_{\rm p}$ remains challenging. Therefore, in the following, we will consider the range $\epsilon_{\rm p}=0.1~\%-0.33~ \%$, for which $\epsilon_{\rm p}$ can be estimated with sufficient accuracy. Within this range, we find slightly larger slopes of $\partial\omega_{\rm G}/\partial\epsilon_{\rm p}=(-57\pm5)$\,cm$^{-1}$/\% strain and $\partial\omega_{\rm 2D}/\partial\epsilon_{\rm p}=(-128\pm10)$\,cm$^{-1}$/\% strain, respectively. These slopes allow us to estimate the Gr\"{u}neisen parameters of the G and 2D modes under biaxial strain, as $\gamma_{\rm G}=\frac{1}{2\omega_{\rm G}^{0}}\frac{\partial \omega_{\rm G}}{\partial \epsilon_{\rm p}}=1.8\pm0.2$ and $\gamma_{\rm 2D}=\frac{1}{2\omega_{\rm 2D}^{0}}\frac{\partial \omega_{\rm 2D}}{\partial \epsilon_{\rm p}}= 2.4\pm 0.2$, respectively, where $\omega_{\rm G}^{0}$ and $\omega_{\rm 2D}^{0}$ are the G- and 2D-mode frequencies in pristine graphene.

\begin{table*} [t!]

\begin{center}
\begin{tabular}{|c|c|c|c|c|c|}
\hline
\hline
 & Method& $\frac{\partial\omega_{\rm 2D}^{}}{\partial\omega_{\rm G}^{}}$ &  $\gamma_{\rm G}^{}$ &  $\gamma_{\rm 2D}^{}$  & $E~(\rm TPa)$\\
\hline
\hline
This work & Raman  & $2.2\pm0.1$ & $1.8\pm0.2$ & $2.4\pm0.2$ &$1.05\pm0.1$ \\
\hline
Graphene bubble \cite{Zabel2011} &Raman + AFM & $2.45\pm 0.3$ & $1.8\pm0.2$ & $2.6\pm0.1$ & $-$\\
\hline
Suspended graphene \cite{Lee2012} &Raman + simulation & $2.2\pm0.2$ & $-$ & $-$ & $2.4\pm0.4$\\
\hline
Pristine graphene \cite{Thomsen2002,Mohiuddin2009,Cheng2011} &First principles & $-$ & $1.8-2.0$ & $2.7$ (ref. \cite{Mohiuddin2009}) & $-$\\
\hline
Suspended graphene \cite{Lee2008} &Nano-indentation & $-$ & $-$ & $-$ & $1.0\pm0.1$\\
\hline
Pristine graphene \cite{Jiang2009,Tan2013} &Molecular dynamics & $-$ & $-$ & $-$ & $1.0\pm0.1$\\
\hline
\end{tabular}
\end{center}
 
    \caption{Comparison of our results with other works. Ratio of the shift rate of the 2D mode to that of the G mode under biaxial tensile strain, Gr\"{u}neisen parameters for the G- and 2D-mode features and Young's modulus of graphene determined in the present study. Our all-optical measurements are compared with experimental and theoretical values reported in the literature.}
\label{table1}
\end{table*}

\section{Determination of the Young's modulus of graphene}

\begin{figure}[!ht]
\begin{center}
\includegraphics[scale=1.0]{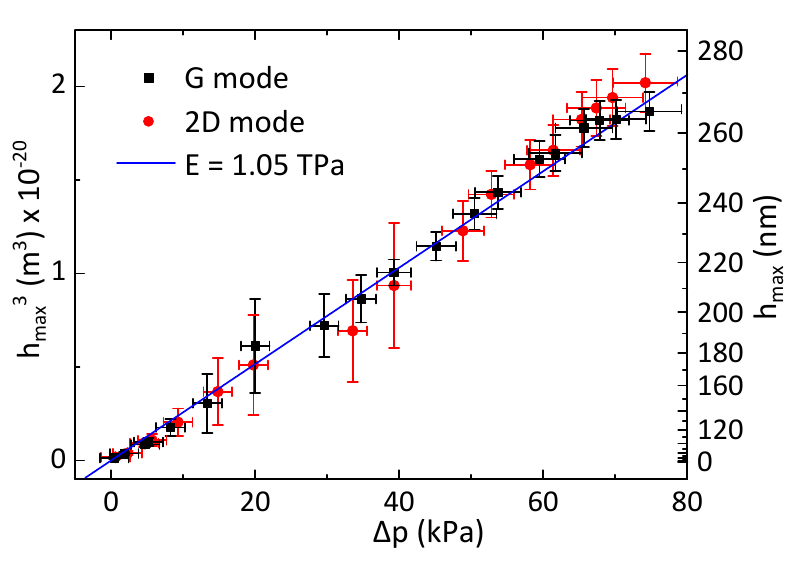}
\caption{Determination of the Young's modulus of graphene. Third power of the height of the graphene blister, $h_{\rm max}$, measured at its center as a function of the pressure load. Data obtained from the measurement of the G- (2D-) mode integrated intensity are shown with black squares (red circles). The straight line is a linear fit, which allows one to deduce a Young's modulus of $E=(1.05~\pm 0.10)~\rm TPa$.}
\label{fig07}
\end{center}
\end{figure}

We now consider the evolution of $h_{\rm max}$, the height measured at the center of the blister, as a function of $\Delta p$. As demonstrated by Hencky in 1915 \cite{Hencky1915,Wan1995}, the third power of deflection of a thin circular plate with negligible bending stiffness, i.e., a membrane, is expected to be proportional to $\Delta p$:

\begin{equation}
\Delta p=\frac{K(\nu)Et}{a^4} h_{\rm max}^3,
\label{eqhvsp}
\end{equation}

where $E$ is the Young's modulus, $t$ is the thickness of the membrane ($t=0.335~\rm nm$ for monolayer graphene), and $K(\nu)$ is a constant that depends on the Poisson ratio of the membrane. In addition, the volume of the blister $V_{\rm B}$ can be written as

\begin{equation}
V_{\rm B}=C(\nu)\pi a^2h_{\rm max},
\end{equation}

where $C(\nu)$ is another constant that is directly related to $K(\nu)$ \cite{Hencky1915,Wan1995}. Similarly to the determination of $\epsilon_{\rm p}$, we also estimate $V_{\rm B}$ for each value of $\Delta p$ (i.e., of $h$), and deduce an average $C(\nu)=0.52\pm0.02$ (see Supplemental Material~\cite{SMnote}). This value is very close to those previously suggested for monolayer graphene, by using $\nu\approx0.16$, the value of bulk graphite \cite{Bunch2008, Koenig2011}. In these conditions, one expects $K(\nu)\approx 3$ \cite{Hencky1915,Wan1995,Bunch2008, Koenig2011}.

In Figure \ref{fig07}, we show the relationship between $h_{\rm max}^3$ and $\Delta p$. Both curves follow very similar linear scalings through the origin, in excellent agreement with Eq. \ref{eqhvsp}. Using $K=3.09$ \cite{Bunch2008, Koenig2011} and $a=(4.1\pm0.1)~\mu$m, we determine the Young's modulus of monolayer graphene as $E=(1.05~\rm \pm 0.1~\rm) TPa$.


\section{Discussion}

A summary of our experimental results and a brief survey of relevant literature values are presented in Table \ref{table1}. Let us first consider the slope $\partial \omega_{\rm 2D}/\partial \omega_{\rm G}$. Our value of $2.2\pm0.1$ is in good agreement with recent studies  by Zabel \textit{et al.} \cite{Zabel2011} on a graphene bubble and by Lee, Yoon, and Cheong \cite{Lee2012} on suspended graphene. Interestingly, we demonstrate that the slope $\partial \omega_{\rm 2D}/\partial \omega_{\rm G}$ is the same at the center of a pressurized blister, where strain is biaxial, and near its edges, where shear deformation (i.e., a uniaxial strain component) is present. We conclude that the value $\partial \omega_{\rm 2D}/\partial \omega_{\rm G}=2.2\pm0.1$, which also has been proposed by Lee \textit{et al.} \cite{Lee2012a} for thermally annealed, supported graphene, seems to be universal for graphene, in the limit of moderate strains below 1\%. As reported, larger uniaxial strains induce shear deformation and subsequent splittings of the Raman features, which strongly depend upon the polarization of the incoming and scattered phonons relative to the crystal orientation \cite{Huang2009,Mohiuddin2009,Huang2010,Frank2011,Yoon2011,Kitt2013}. These factors complicate the determination of $\partial \omega_{\rm 2D}/\partial \omega_{\rm G}$ and consequently of the Gr\"{u}neisen parameters. 

Under biaxial strain, the Gr\"{u}neisen parameters are determined more reliably, since these are simply proportional to $\partial \omega_{\rm 2D}/\partial \omega_{\rm G}$. The main challenge is then to determine the amount of strain with accuracy. Remarkably, our all-optical determination of $\epsilon_{\rm p}$, $\gamma_{\rm G}$ and $\gamma_{\rm 2D}$ agrees well with an estimation based on combined AFM and Raman measurements on a graphene bubble on a Si/SiO$_2$ substrate \cite{Zabel2011}. We also find good agreement with theoretical predictions \cite{Thomsen2002,Mohiuddin2009,Cheng2011}. Interestingly, we demonstrate that a direct determination of $\epsilon_{\rm p}$ from the integrated intensity of the Raman features can be performed \textit{in situ}, as a function of $\Delta p$. 
Although the lateral resolution of our approach is set by the diffraction limit, the measured heights can be estimated with precisions up to about $5~\rm nm$. Our approach also has the major advantage of being contactless and minimally invasive, as opposed to scanning probe techniques, such as AFM, where sample-tip interaction is known to lead to artifacts when probing the topography of the suspended membrane. In addition, our experimental setup is obviously easier and cheaper to implement than an \textit{in situ} AFM setup, which would be an alternative way to probe the blister topography, as a function of a controllable pressure load with a better lateral resolution. In any event, we note that precise determinations of the Gr\"{u}neisen parameters of graphene remain difficult, since these typically combine a local Raman measurement with an estimation of the amount of strain that is averaged over a much larger area. These experimental difficulties may, in part, explain the relatively large spread in the experimental values of $\gamma_{\rm G}$ and $\gamma_{\rm 2D}$ reported in the literature.

Finally, our measurement of the Young's modulus of graphene matches the value of bulk graphite and is in excellent agreement with values obtained by using scanning probe techniques, such as nanoindentation \cite{Lee2008} and AFM \cite{Koenig2011}, as well as with molecular dynamics simulations \cite{Jiang2009,Tan2013}. Here, the Young's modulus is determined with accuracy using a simple, all-optical and minimally invasive approach. We note, that Lee, Yoon, and Cheong have recently proposed a significantly larger value of $E$ (see Table \ref{table1} and Ref. \cite{Lee2012}). The latter estimate is obtained from a comparison of Raman scattering measurements with finite elements simulations \cite{Lee2012}. We believe that this discrepancy is due to the fact that $\epsilon_{\rm p}$ has been qualitatively estimated using previously reported Raman measurements on uniaxially strained supported graphene \cite{Yoon2011}. This difference further highlights the interest of our approach, which allows a combined study of the topography and of the vibrational properties of suspended graphene, from a consistent set of measurements.


\section{Conclusion}
 Using micro-Raman scattering spectroscopy, we have performed a constant $N$ blister test on a suspended graphene membrane under a uniform pressure load. By analyzing the frequencies and the integrated intensities of the main Raman features of graphene, we reconstruct the blister topography, and deduce the Gr\"{u}neisen parameters and the Young's modulus. Our analysis reveals that the intensity of the Raman features of a suspended graphene membrane can vary by one order of magnitude for pressure changes of only a few kPa [see also Fig. \ref{fig04}(a)]. Considering Eq. \ref{eqhvsp}, the relative change in blister height will be particularly strong close to pressure equilibrium, i.e., for $\Delta p\approx 0 ~\rm kPa$. This result suggests that typical fluctuations of the atmospheric pressure, as low as $1~\rm kPa$, could be sensed with accuracy, by using \textit{graphene-based barometers}. Our approach can be implemented for all-optical adhesion studies, under larger pressure loads \cite{Koenig2011} and could directly be generalized to other two-dimensional materials, such as transition metal dichalcogenides~\cite{Castellanos-Gomez2013}.
 
More generally, we hope that our study may inspire original research efforts at the interface between electromechanics~\cite{Ekinci2005}, optomechanics~\cite{Robert2012} and mechanical engineering. Indeed, the  mechanical response of micro- and nanoresonators is highly sensitive to the stress conditions~\cite{Southworth2009,Singh2010,Chen2009,Mohanty2002,Bercu2009,Bauer2008,Reserbat2012}. Our approach should permit to measure the topography and strain distribution (including built-in strain~\cite{Metten2013, StarmanJr2003}) of micro- and nanoresonators, in various environments, provided an interference pattern can develop. For instance, a readout of the mechanical resonance of few-layer graphene cantilevers through the frequency of the Raman features has recently been demonstrated ~\cite{Reserbat2012}. A stimulating challenge is now to implement a real-time readout~\cite{Xue2007,Pomeroy2008} of the mechanical resonances of nanosystems based on Raman scattering spectroscopy.
 

\paragraph*{\textbf{Acknowledgement}}
We are grateful to Antoine Reserbat-Plantey and Guillaume Froehlicher for fruitful discussions. We thank Romain Bernard, Sabine Siegwald and Hicham Majjad for help with sample fabrication in the STNano clean room facility, and Bernard Doudin, Fabien Chevrier and Arnaud Boulard for technical support. We acknowledge financial support from the Centre National de la Recherche Scientifique (CNRS), Universit\'e de Strasbourg, C'Nano GE and the Agence Nationale de Recherche (ANR) under Grant No.~QuandDoGra 12 JS10-00101.






%


\onecolumngrid
\newpage
\begin{center}
{\Large\textbf{Supplemental Material}}
\end{center}

\setcounter{equation}{0}
\setcounter{figure}{0}
\setcounter{section}{0}
\renewcommand{\theequation}{S\arabic{equation}}
\renewcommand{\thefigure}{S\arabic{figure}}
\renewcommand{\thesection}{S\arabic{section}}



\section{Determination of the pressure load}
\label{pressure}
A key point in our analysis is the calibration of the pressure load $\Delta p$. This measurement is not straightforward since the pit depth $h_0$ is on the same order of magnitude as the blister height $h_{\rm max}$. We have therefore measured the blister topography (see main text) and deduced the blister volume $V_{\rm B}$ at $p_{\rm ext}\approx10^{-2}~\rm Pa$, starting from $p_{\rm int}=(100\pm2) ~\rm kPa$. Bulging of the graphene provides a larger volume ($V_{\rm B}+V_{\rm 0}$, where $V_{\rm 0}$ is the volume of the cylindrical pit) for the air molecules trapped under the graphene blister. From the ideal gas law, we can then deduce the reduced inner pressure $p_{\rm int}=(74\pm 5)~ \rm kPa$, considering the error bars in $V_{\rm 0}$ and in the atmospheric pressure. We then assume that in the limit low pressure loads $\Delta p=p_{\rm int}-p_{\rm ext}$, the downshifts of the Raman features scale linearly with $\Delta p$~\cite{Zabel2011}. Considering the two limiting cases of $p_{ext}\approx~10^{-2}~\rm Pa$ and $p_{\rm ext}= 100~\rm kPa$, we thus get an estimation of $\Delta p$ with an error bar of a few kPa for each measurement. We note that very similar values of $\Delta p$ are obtained from the downshifts of the G and 2D mode features.

\section{Determination of the blister height from the 2D mode intensity}
\label{2Dmode}

\begin{figure}[!htb]
\begin{center}
\includegraphics[scale=1]{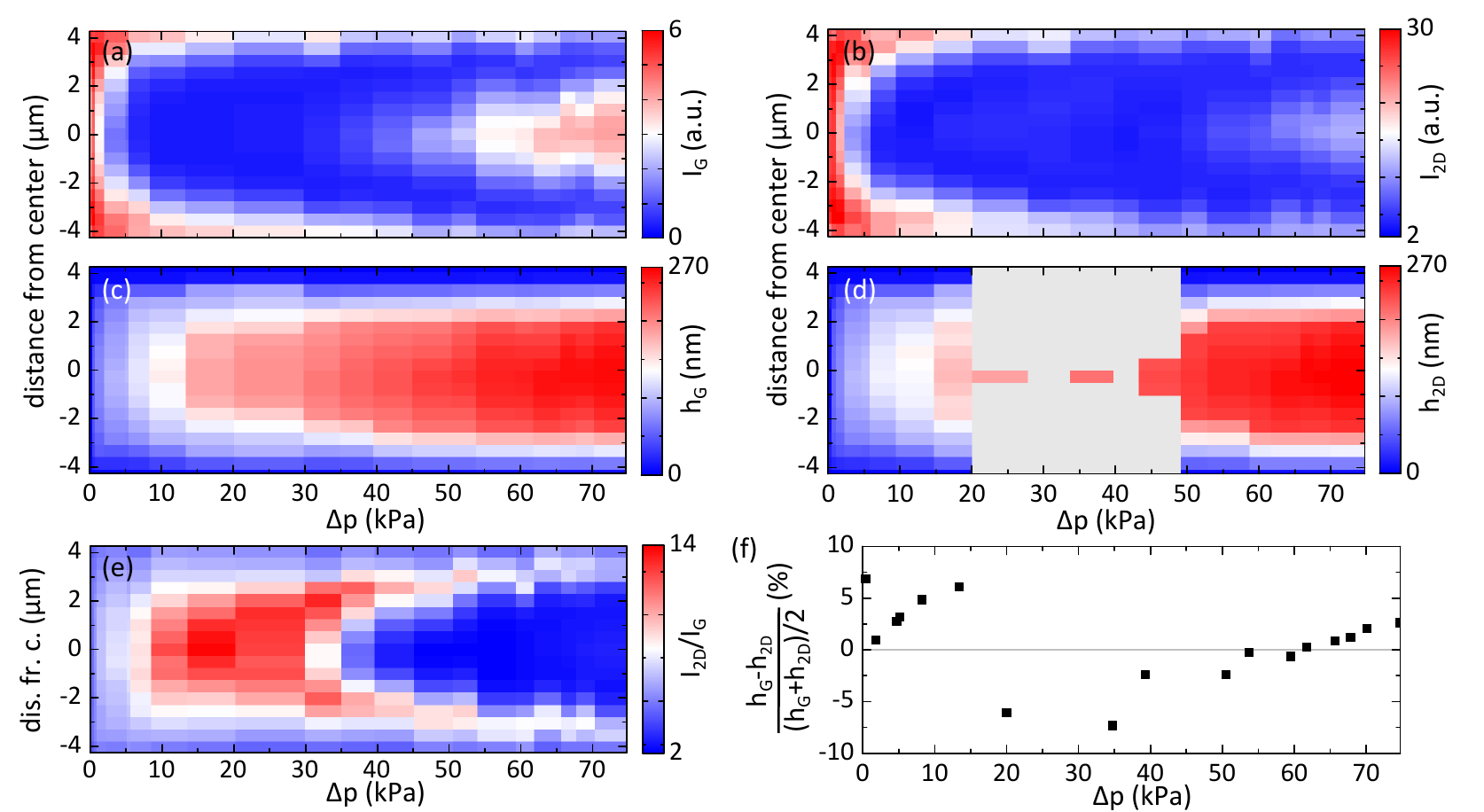}
\caption{a) Raman G mode intensity $I_{\rm G}^{}$, b) Raman 2D mode intensity $I_{\rm 2D}^{}$ (b), c) blister height deduced from $I_{\rm G}^{}$, d) blister height deduced from $I_{\rm 2D}^{}$, e) integrated intensity ratio $I_{\rm 2D}^{}/I_{\rm G}^{}$, measured on sample A, as a function of the distance from the blister center $r$ and the pressure load $\Delta p$. The area in light gray in d) corresponds to a regime of low Raman intensity, where the blister topography cannot be determined accurately. f) Relative difference between the maximal blister heights deduced using the G and 2D mode intensities.}
\label{SI01}
\end{center}
\end{figure}

As discussed in the main manuscript, the blister height can be deduced from the integrated intensity of the G and 2D mode features. Figure~\ref{SI01} shows the blister height deduced using the result of the interference model by Yoon \textit{et al.}~\cite{Yoon2009}. The heights $h_{\rm G}$ and $h_{\rm 2D}$, obtained using the integrated intensity of the G and 2D modes are very similar. The relative difference between $h_{\rm G}$ and $h_{\rm 2D}$ is plotted in figure Figure~\ref{SI01} and does not exceed $10~\%$.

\newpage

\section{Influence of the focusing conditions}
\label{focus}
\begin{figure}[!htb]
\begin{center}
\includegraphics[scale=1.0]{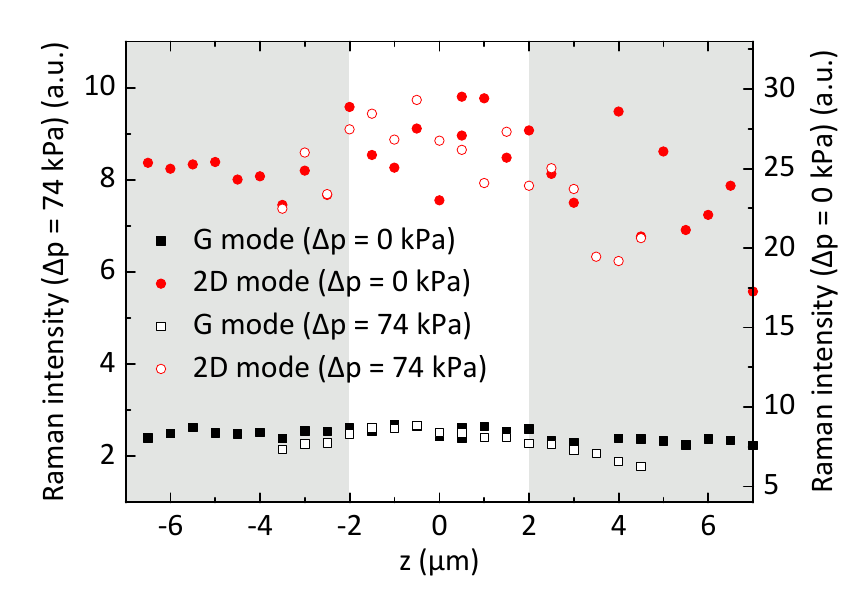}
\caption{Dependence of $I_{\rm G}$ (squares) and $I_{\rm 2D}$ (circles) on the laser beam waist axial position, denoted $z$. Data taken at $\Delta p=0~\rm kPa$ and $\Delta p=74~\rm kPa$ are shown with full and open symbols, respectively.}
\label{SI02}
\end{center}
\end{figure}

Figure~\ref{SI02} shows the variations of $I_{\rm G}^{}$ and $I_{\rm 2D}^{}$ as a function of the axial position of the laser beam waist relative to the graphene membrane, for two extreme values of $\Delta p=74~\rm kPa$ and $\Delta p=0~\rm kPa$. During a measurement run, we estimate that the laser focus varies by less than $\pm 1\rm \mu m$. Within a range of $\pm2~\rm \mu m$, no appreciable changes in the Raman intensities could be observed. We therefore conclude that our results are not affected by the focusing conditions and that the large changes in the Raman intensity reported here are only due to optical interference effects.

\section{Polarization resolved measurements}
\label{polar}
\begin{figure}[!htb]
\begin{center}
\includegraphics[scale=1.0]{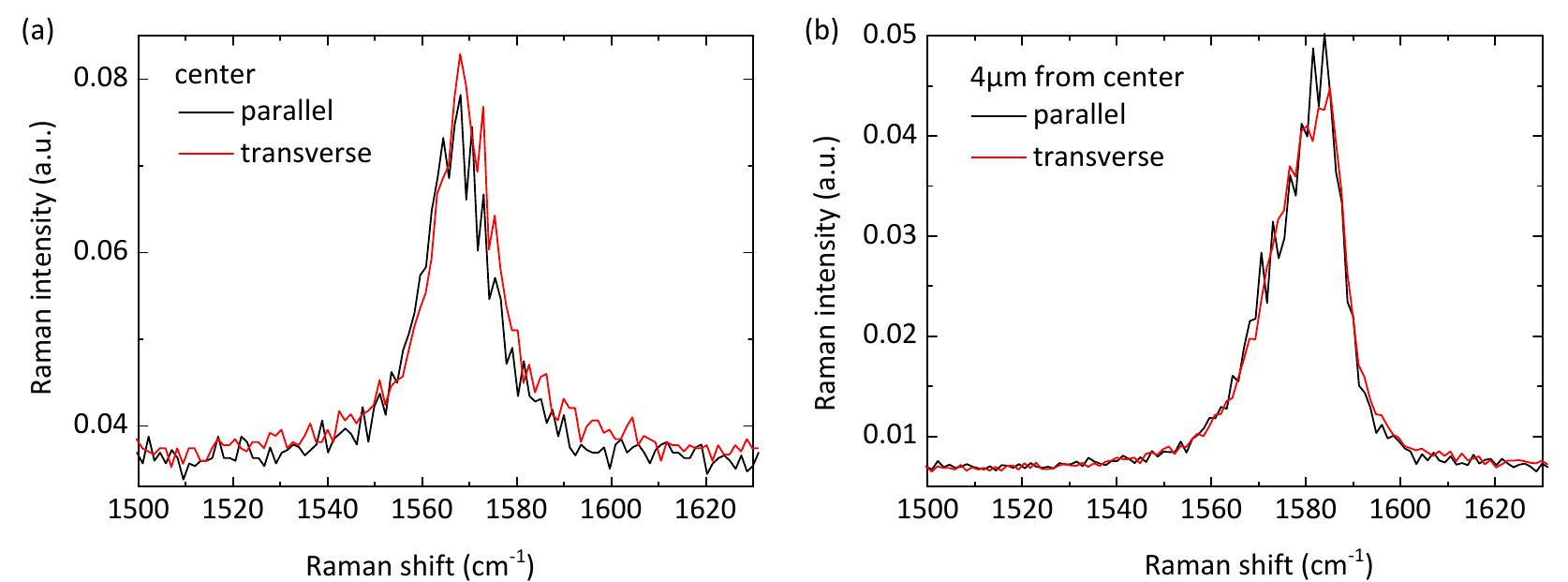}
\caption{Raman G mode spectra recorded at $\Delta p=74~\rm kPa$ at the center of the blister (a) and at $4~\rm \mu m$ from the center (b).}
\label{SI03}
\end{center}
\end{figure}

Figure~\ref{SI03} shows Raman G mode spectra, measured on sample A at $\Delta p=74~\rm kPa$, at the center of the blister and at $r=4~\rm \mu m$ from the center.  The spectra recorded using parallel and perpendicular polarizations for the laser beam and the analyser are very similar. This demonstrates that no significant signature of uniaxial strain can be measured here. As discussed in the main text, the bimodal lineshape observed at $r=4~\rm \mu m$ is assigned to the superposition of the Raman responses of the supported and suspended regions, due to the finite size of the laser spot ($ 1.2~\rm \mu m$ full width at half maximum).

\newpage

\section{Data measured on other samples}
\label{other}
We have measured on three different suspended samples, with a same value of $a\approx4~\rm \mu m$ but with different values of $h_0$. The main manuscript discussed the results obtained on sample A $(h_0=(395\pm10)~\rm nm)$, and in the following we also present data on samples B and C, both of which with a pit depth $(h_0=(340\pm10)~\rm nm)$. Samples B and C were left under vacuum prior to measurements, so that they could undergo significant leakage.

\subsection{Correlation between the 2D and G mode frequencies.} 
\begin{figure}[!htb]
\begin{center}
\includegraphics[scale=1.0]{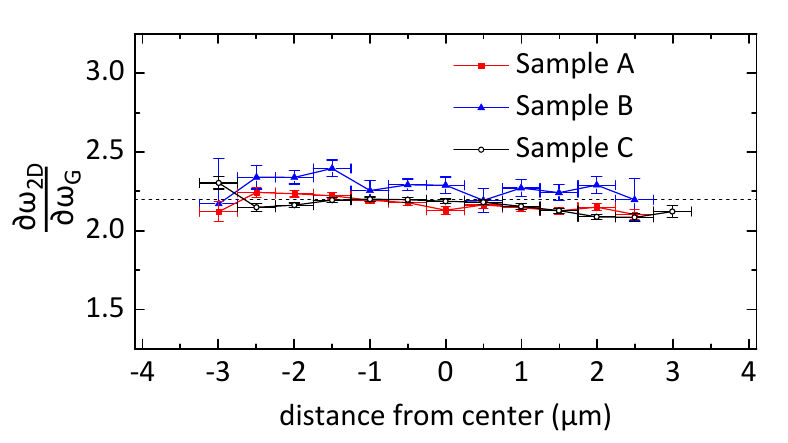}
\caption{ $\partial \omega_{\rm 2D}^{}/\partial \omega_G$ as a function of $r$, the distance from the blister center, for samples A, B, C.}
\label{SI04}
\end{center}
\end{figure}

Figure \ref{SI04} shows data similar to Figure 2e of the main manuscript. The slope $\partial \omega_{\rm 2D}^{}/\partial \omega_{\rm G}^{}$ is systematically close to 2.2, confirming the measurements on sample A.

\subsection{Determination of $\Delta p$ for samples B and C.} In the main text, we assumed that the number of gas molecules under the suspended graphene in sample A was constant, because the values of $I_{\rm G}$, $I_{\rm 2D}$, $\omega_{\rm G}^{}$ and $\omega_{\rm 2D}^{}$ at atmospheric pressure are the same before and after the measurement series as a function of $\Delta p$. Since the initial number of trapped molecules is not known for samples B and C,  it is not possible to directly deduce a value for $p_{\rm int}$ at $p_{\rm ext}\approx10^{-2}\rm~Pa$, using the ideal gas law. In this case, we estimate $\Delta p$ by making the following assumptions:
 
\noindent i) the maximum value of $\omega_G$ during a measurement run as a function of $\Delta p$ corresponds to unstrained flat graphene.  At this particular value, a nearly flat profile is found for $I_{\rm G}^{}$ and $I_{\rm 2D}^{}$ over the blister.

\noindent ii) $\omega_{\rm G}^{},~\omega_{\rm 2D}^{} \propto  \Delta p$. We then used the linear relationship between $\Delta p$ and the Raman frequencies deduced from the measurements on sample A.

As a consequence,the pressure load $\Delta p$ in samples B and C is determined with similar uncertainty but with a potentially greater systematic error than for sample A. 


\subsection{Evolution of $I_{\rm G}^{}$ and $I_{\rm 2D}^{}$ as a function of $\Delta p$ for sample B.} As discussed above, sample B  ($h_0=(340\pm10)~\rm nm$) was left under high vacuum for more than one day. Over this period, we observed a reduction of the blister height, and a reduced softening of the Raman features, due to slow diffusion of the trapped air molecules through the substrate. \textbf{In these conditions, increasing $p_{\rm ext}$ up to ambient pressure results in $\Delta p<0$ and thus, a concave membrane}.

\begin{figure}[!htb]
\begin{center}
\includegraphics[scale=1.0]{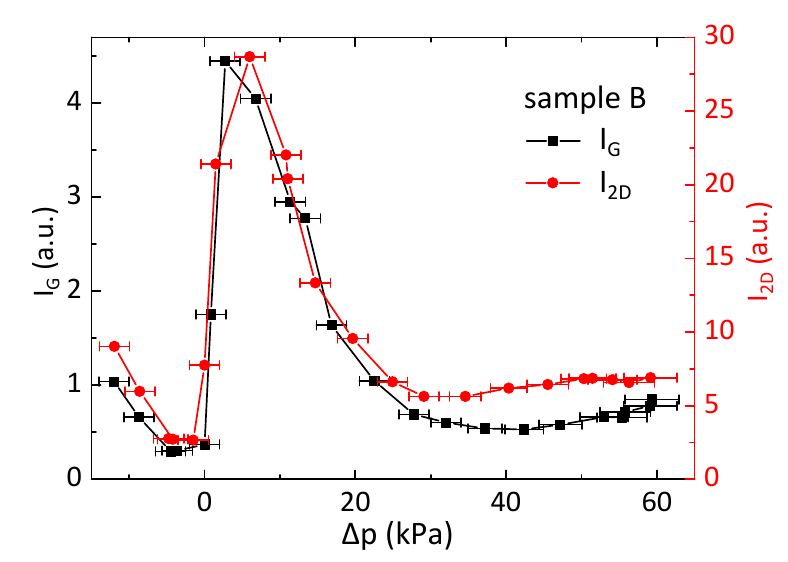}
\caption{ Evolution of the integrated intensities of the G and 2D mode features measured at the center of sample B, as a function of the pressure load $\Delta p$.}
\label{SI05}
\end{center}
\end{figure}

\begin{figure}[!hbt]
\begin{center}
\includegraphics[scale=1.0]{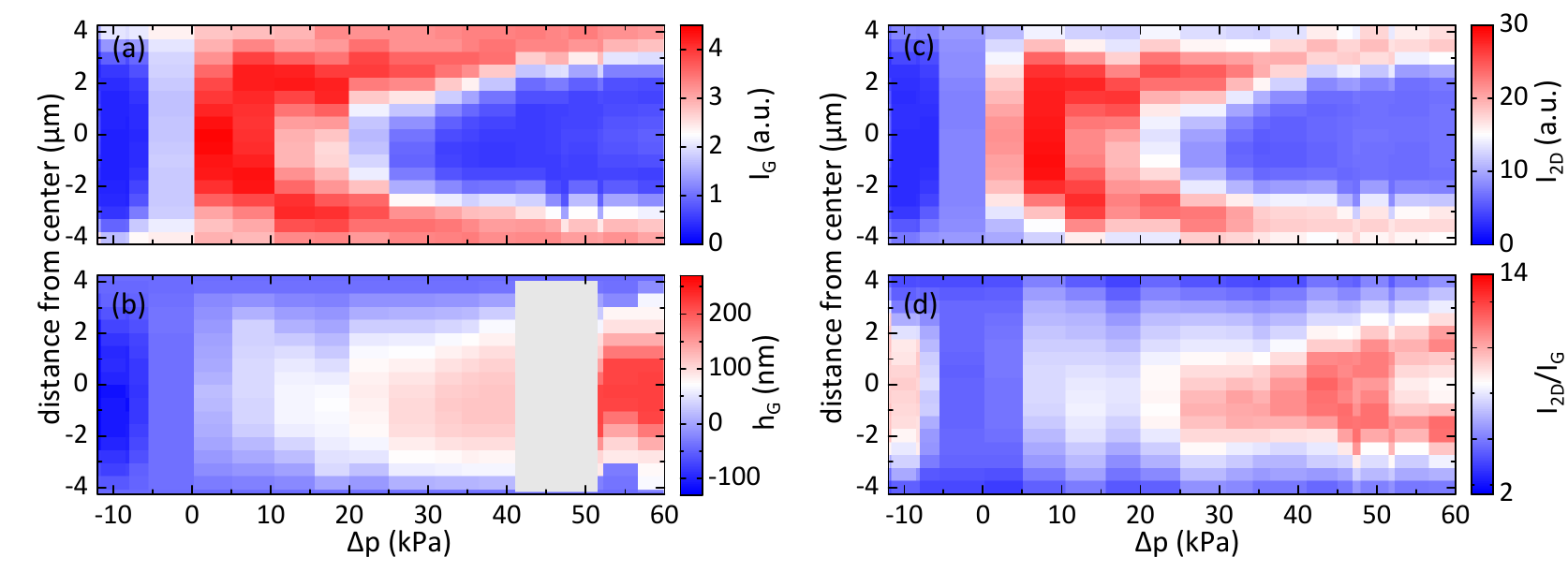}
\caption{ a) Raman G mode intensity $I_{\rm G}^{}$, b) blister height deduced from $I_{\rm G}^{}$, c) Raman 2D mode intensity $I_{\rm 2D}^{}$ d) integrated intensity ratio $I_{\rm 2D}^{}/I_{\rm G}^{}$, plotted as a function of the distance from the blister center $r$ and the pressure load $\Delta p$. The area in light gray in c) corresponds to a regime of low Raman intensity, where the blister topography cannot be determined accurately.}
\label{SI06}
\end{center}
\end{figure}

 In Figure \ref{SI05}, we present the evolution of $I_{\rm G}^{}$ and $I_{\rm 2D}^{}$ as in Figure 3a of the main manuscript. The data in Figure \ref{SI05} can also be well understood considering Raman enhancement effects~\cite{Yoon2009}. Since $h_0=(340\pm10)~\rm nm$ is smaller than in sample A, for which $h_0=(395\pm10)~\rm nm$, we observe an overall shift of the interference pattern relative to the data in Figure 3a of the main manuscript.  In particular at $\Delta p\approx0~\rm kPa$, $I_{\rm G}^{}$ and $I_{\rm 2D}^{}$ are not as close to a maximum (expected at $h_{\rm tot}\approx 420~\rm nm $) as in the case of sample A, and the maximum of $I_{\rm G}^{}$ and $I_{\rm 2D}^{}$, corresponding to $h_{\rm tot}\approx 420~\rm nm $ is reached at larger $\Delta p$ than for sample A. Finally, a minimum in $I_{\rm G}^{}$ and $I_{\rm 2D}^{}$ is observed at negative $\Delta p$ and is assigned to $h_{\rm tot}\approx 270~\rm nm $, \textit{i.e.,} a deflection of the graphene membrane of $\approx 70~\rm nm$ towards the Si substrate. We have attained downwards deflections of as much as $\approx 120~\rm nm$ in sample B.  Altogether, the behavior observed in Figure \ref{SI05} bolsters our analysis of the blister topography based on optical interference effects on the Raman intensity. Data similar to Figure \ref{SI01} recorded on sample B are shown in Figure \ref{SI06}.

We note, that in the ranges $-10~\rm kPa<\Delta p<0~\rm kPa$ and $\Delta p>20~\rm kPa$, $I_{\rm G}^{}$ and $I_{\rm 2D}^{}$ are close to a minimum. This means that in these cases, the heights $h_{\rm max}$ are obtained with lower accuracy than for sample A, where the larger value of $h_{0}=(395\pm10)~\rm nm$ permits a much more precise determination of $h_{max}$ for $\Delta p>20~\rm kPa$. 



\subsection{Correlation between the Raman frequencies and the estimated strain} 
\begin{figure}[!htb]
\begin{center}
\includegraphics[scale=1.0]{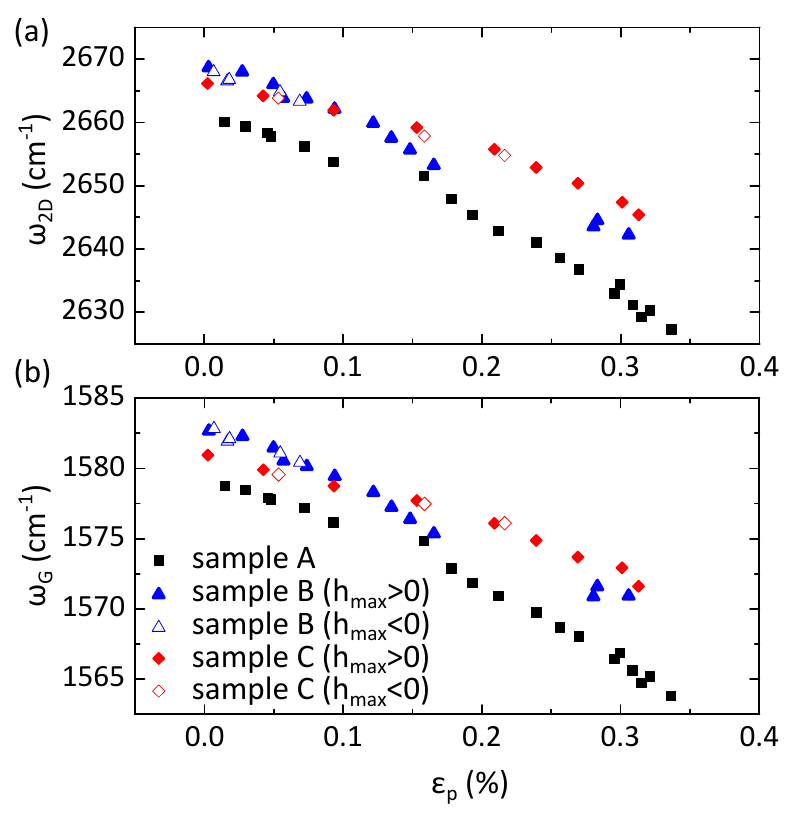}
\caption{Evolution of the G mode (a)  and 2D mode (b) frequencies measured at the center of samples A, B, C, as a function of the tensile strain $\epsilon_{\rm p}$ induced by the uniform pressure load. For clarity, the error bars have been omitted. Full (open) symbols represent measurements at positive (negative) pressure load.}
\label{SI07}
\end{center}
\end{figure}

Figure \ref{SI07} compares the evolution of $\omega_{\rm G}^{}$ and $\omega_{\rm 2D}^{}$ as a function of $\epsilon_{\rm p}$  for samples A, B, C.
We also find that the Raman frequencies decrease roughly linearly with $\epsilon_{\rm p}$ and that the slopes are quite similar to the values observed in sample A. We note, however that the Raman frequencies measured on samples B and C are slightly upshifted with respect to the data in sample A. In particular, when $\epsilon_{\rm p}$ vanishes, we find $\omega_{\rm G}^{} \approx 1581~\rm cm^{-1}$ and $\omega_{\rm 2D}^{}\approx 2666~\rm cm^{-1}$, the expected values for undoped, unstrained graphene (for $\lambda_{\rm laser}=532~\rm nm$). This suggests that samples B and C experience negligible built-in strain~\cite{Metten2013}.

\newpage

\subsection{Correlation between the blister height and the pressure load.} 
\begin{figure}[!htb]
\begin{center}
\includegraphics[scale=0.95]{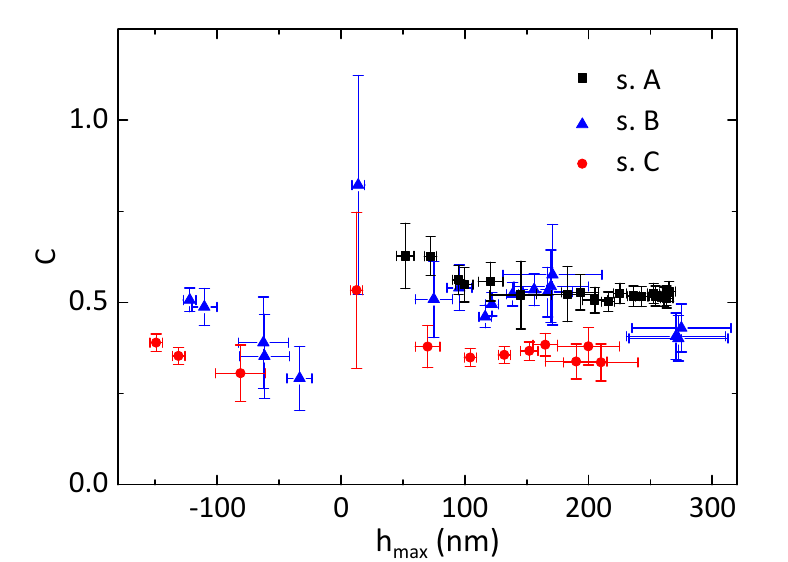}
\caption{$C$ vs. $h_{\rm max}$ for samples A, B, C. The volume $V_{\rm B}$ is determined by integrating the measured height profile.}
\label{SI08}
\end{center}
\end{figure}

As discussed in the main text, the Young'smodulus can be estimated from the evolution of $h$ as a function of  $\Delta p$, through Eq. 1, provided the coefficient $K(\nu)$ is known. In Figure \ref{SI08}, we show the values of $C(\nu)=\frac{V_{\rm B}}{\pi a^2 h_{\rm max}}$, as introduced in Eq. 2. The values of $C(\nu)$, determined with varying $\Delta p$ are close to 0.5 for all samples. This is consistent with the value of $K(\nu)=3.09$ that we have used in our analysis~\cite{Hencky1915,Wan1995,Bunch2008,Koenig2011}.

\begin{figure}[!hbt]
\begin{center}
\includegraphics[scale=0.95]{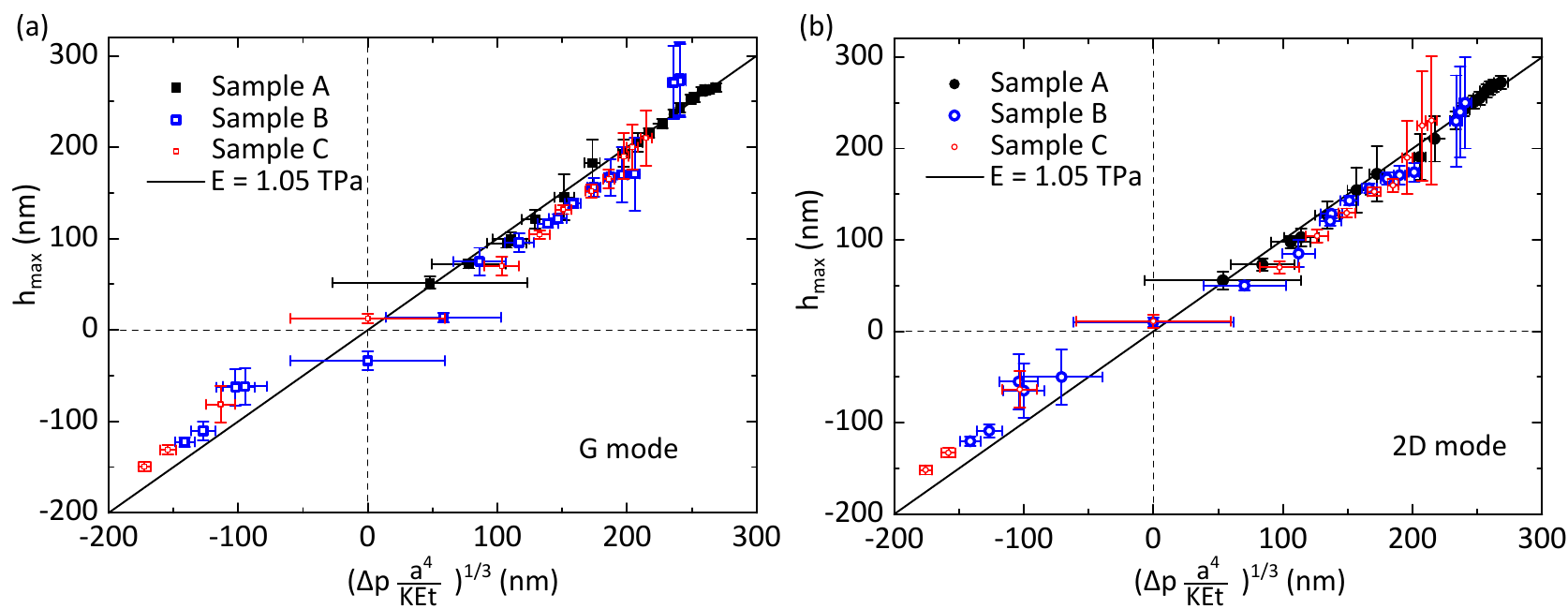}
\caption{Height of the graphene blister, $h_{\rm max}$, measured at its center represented as a function of $\left(\frac{\Delta p\:a^4}{K(\nu)Et}\right)^{1/3}$. Data obtained from the measurement of the G (2D) mode integrated intensity are shown in a) (b) for samples A, B, C. The straight line corresponds to a Young's modulus of $E=(1.05~\rm \pm 0.10)~\rm TPa$.}
\label{SI09}
\end{center}
\end{figure}

Finally, figure \ref{SI09} compares the evolution of $h_{\rm max}$ as a function of $\left(\frac{\Delta p\:a^4}{K(\nu)Et}\right)^{1/3}$  for samples A, B and C. In spite of the greater uncertainty in the determination of $h_{\rm max}$ and $\Delta p^{1/3}$ for  samples B and C, we find that the data measured on these samples is in good agreement with the data measured on sample A, for which we estimated $E=1.05\pm0.1~\rm TPa$.

\newpage


\begin{thebibliography}{56}%
\makeatletter
\providecommand \@ifxundefined [1]{%
 \@ifx{#1\undefined}
}%
\providecommand \@ifnum [1]{%
 \ifnum #1\expandafter \@firstoftwo
 \else \expandafter \@secondoftwo
 \fi
}%
\providecommand \@ifx [1]{%
 \ifx #1\expandafter \@firstoftwo
 \else \expandafter \@secondoftwo
 \fi
}%
\providecommand \natexlab [1]{#1}%
\providecommand \enquote  [1]{``#1''}%
\providecommand \bibnamefont  [1]{#1}%
\providecommand \bibfnamefont [1]{#1}%
\providecommand \citenamefont [1]{#1}%
\providecommand \href@noop [0]{\@secondoftwo}%
\providecommand \href [0]{\begingroup \@sanitize@url \@href}%
\providecommand \@href[1]{\@@startlink{#1}\@@href}%
\providecommand \@@href[1]{\endgroup#1\@@endlink}%
\providecommand \@sanitize@url [0]{\catcode `\\12\catcode `\$12\catcode
  `\&12\catcode `\#12\catcode `\^12\catcode `\_12\catcode `\%12\relax}%
\providecommand \@@startlink[1]{}%
\providecommand \@@endlink[0]{}%
\providecommand \url  [0]{\begingroup\@sanitize@url \@url }%
\providecommand \@url [1]{\endgroup\@href {#1}{\urlprefix }}%
\providecommand \urlprefix  [0]{URL }%
\providecommand \Eprint [0]{\href }%
\providecommand \doibase [0]{http://dx.doi.org/}%
\providecommand \selectlanguage [0]{\@gobble}%
\providecommand \bibinfo  [0]{\@secondoftwo}%
\providecommand \bibfield  [0]{\@secondoftwo}%
\providecommand \translation [1]{[#1]}%
\providecommand \BibitemOpen [0]{}%
\providecommand \bibitemStop [0]{}%
\providecommand \bibitemNoStop [0]{.\EOS\space}%
\providecommand \EOS [0]{\spacefactor3000\relax}%
\providecommand \BibitemShut  [1]{\csname bibitem#1\endcsname}%
\let\auto@bib@innerbib\@empty
\bibitem [{\citenamefont {Novoselov}\ \emph {et~al.}(2005)\citenamefont
  {Novoselov}, \citenamefont {Jiang}, \citenamefont {Schedin}, \citenamefont
  {Booth}, \citenamefont {Khotkevich}, \citenamefont {Morozov},\ and\
  \citenamefont {Geim}}]{Novoselov2005}%
  \BibitemOpen
  \bibfield  {author} {\bibinfo {author} {\bibfnamefont {K.S.}\ \bibnamefont
  {Novoselov}}, \bibinfo {author} {\bibfnamefont {D.}~\bibnamefont {Jiang}},
  \bibinfo {author} {\bibfnamefont {F.}~\bibnamefont {Schedin}}, \bibinfo
  {author} {\bibfnamefont {T.J.}\ \bibnamefont {Booth}}, \bibinfo {author}
  {\bibfnamefont {V.V.}\ \bibnamefont {Khotkevich}}, \bibinfo {author}
  {\bibfnamefont {S.V.}\ \bibnamefont {Morozov}}, \ and\ \bibinfo {author}
  {\bibfnamefont {A.K.}\ \bibnamefont {Geim}},\ }\bibfield  {title} {\enquote
  {\bibinfo {title} {Two-dimensional atomic crystals},}\ }\href
  {http://www.pnas.org/content/102/30/10451.abstract} {\bibfield  {journal}
  {\bibinfo  {journal} {Proceedings of the National Academy of Sciences of the
  United States of America}\ }\textbf {\bibinfo {volume} {102}},\ \bibinfo
  {pages} {10451--10453} (\bibinfo {year} {2005})}\BibitemShut {NoStop}%
\bibitem [{\citenamefont {Castro~Neto}\ \emph {et~al.}(2009)\citenamefont
  {Castro~Neto}, \citenamefont {Guinea}, \citenamefont {Peres}, \citenamefont
  {Novoselov},\ and\ \citenamefont {Geim}}]{Castroneto2009}%
  \BibitemOpen
  \bibfield  {author} {\bibinfo {author} {\bibfnamefont {A.~H.}\ \bibnamefont
  {Castro~Neto}}, \bibinfo {author} {\bibfnamefont {F.}~\bibnamefont {Guinea}},
  \bibinfo {author} {\bibfnamefont {N.~M.~R.}\ \bibnamefont {Peres}}, \bibinfo
  {author} {\bibfnamefont {K.~S.}\ \bibnamefont {Novoselov}}, \ and\ \bibinfo
  {author} {\bibfnamefont {A.~K.}\ \bibnamefont {Geim}},\ }\bibfield  {title}
  {\enquote {\bibinfo {title} {The electronic properties of graphene},}\ }\href
  {\doibase 10.1103/RevModPhys.81.109} {\bibfield  {journal} {\bibinfo
  {journal} {Reviews of Modern Physics}\ }\textbf {\bibinfo {volume} {81}},\
  \bibinfo {pages} {109} (\bibinfo {year} {2009})}\BibitemShut {NoStop}%
\bibitem [{\citenamefont {Lee}\ \emph {et~al.}(2008)\citenamefont {Lee},
  \citenamefont {Wei}, \citenamefont {Kysar},\ and\ \citenamefont
  {Hone}}]{Lee2008}%
  \BibitemOpen
  \bibfield  {author} {\bibinfo {author} {\bibfnamefont {Changgu}\ \bibnamefont
  {Lee}}, \bibinfo {author} {\bibfnamefont {Xiaoding}\ \bibnamefont {Wei}},
  \bibinfo {author} {\bibfnamefont {Jeffrey~W.}\ \bibnamefont {Kysar}}, \ and\
  \bibinfo {author} {\bibfnamefont {James}\ \bibnamefont {Hone}},\ }\bibfield
  {title} {\enquote {\bibinfo {title} {Measurement of the elastic properties
  and intrinsic strength of monolayer graphene},}\ }\href {\doibase
  10.1126/science.1157996} {\bibfield  {journal} {\bibinfo  {journal}
  {Science}\ }\textbf {\bibinfo {volume} {321}},\ \bibinfo {pages} {385--388}
  (\bibinfo {year} {2008})}\BibitemShut {NoStop}%
\bibitem [{\citenamefont {Koenig}\ \emph {et~al.}(2011)\citenamefont {Koenig},
  \citenamefont {Boddeti}, \citenamefont {Dunn},\ and\ \citenamefont
  {Bunch}}]{Koenig2011}%
  \BibitemOpen
  \bibfield  {author} {\bibinfo {author} {\bibfnamefont {Steven~P.}\
  \bibnamefont {Koenig}}, \bibinfo {author} {\bibfnamefont {Narasimha~G.}\
  \bibnamefont {Boddeti}}, \bibinfo {author} {\bibfnamefont {Martin~L.}\
  \bibnamefont {Dunn}}, \ and\ \bibinfo {author} {\bibfnamefont {J.~Scott}\
  \bibnamefont {Bunch}},\ }\bibfield  {title} {\enquote {\bibinfo {title}
  {Ultrastrong adhesion of graphene membranes},}\ }\href
  {http://dx.doi.org/10.1038/nnano.2011.123} {\bibfield  {journal} {\bibinfo
  {journal} {Nat Nano}\ }\textbf {\bibinfo {volume} {6}},\ \bibinfo {pages}
  {543--546} (\bibinfo {year} {2011})}\BibitemShut {NoStop}%
\bibitem [{\citenamefont {Bunch}\ \emph {et~al.}(2008)\citenamefont {Bunch},
  \citenamefont {Verbridge}, \citenamefont {Alden}, \citenamefont {van~der
  Zande}, \citenamefont {Parpia}, \citenamefont {Craighead},\ and\
  \citenamefont {McEuen}}]{Bunch2008}%
  \BibitemOpen
  \bibfield  {author} {\bibinfo {author} {\bibfnamefont {J.~Scott}\
  \bibnamefont {Bunch}}, \bibinfo {author} {\bibfnamefont {Scott~S.}\
  \bibnamefont {Verbridge}}, \bibinfo {author} {\bibfnamefont {Jonathan~S.}\
  \bibnamefont {Alden}}, \bibinfo {author} {\bibfnamefont {Arend~M.}\
  \bibnamefont {van~der Zande}}, \bibinfo {author} {\bibfnamefont {Jeevak~M.}\
  \bibnamefont {Parpia}}, \bibinfo {author} {\bibfnamefont {Harold~G.}\
  \bibnamefont {Craighead}}, \ and\ \bibinfo {author} {\bibfnamefont {Paul~L.}\
  \bibnamefont {McEuen}},\ }\bibfield  {title} {\enquote {\bibinfo {title}
  {Impermeable atomic membranes from graphene sheets},}\ }\href {\doibase
  10.1021/nl801457b} {\bibfield  {journal} {\bibinfo  {journal} {Nano Lett.}\
  }\textbf {\bibinfo {volume} {8}},\ \bibinfo {pages} {2458--2462} (\bibinfo
  {year} {2008})}\BibitemShut {NoStop}%
\bibitem [{\citenamefont {Novoselov}\ \emph {et~al.}(2004)\citenamefont
  {Novoselov}, \citenamefont {Geim}, \citenamefont {Morozov}, \citenamefont
  {Jiang}, \citenamefont {Zhang}, \citenamefont {Dubonos}, \citenamefont
  {Grigorieva},\ and\ \citenamefont {Firsov}}]{Novoselov2004}%
  \BibitemOpen
  \bibfield  {author} {\bibinfo {author} {\bibfnamefont {K.~S.}\ \bibnamefont
  {Novoselov}}, \bibinfo {author} {\bibfnamefont {A.~K.}\ \bibnamefont {Geim}},
  \bibinfo {author} {\bibfnamefont {S.~V.}\ \bibnamefont {Morozov}}, \bibinfo
  {author} {\bibfnamefont {D.}~\bibnamefont {Jiang}}, \bibinfo {author}
  {\bibfnamefont {Y.}~\bibnamefont {Zhang}}, \bibinfo {author} {\bibfnamefont
  {S.~V.}\ \bibnamefont {Dubonos}}, \bibinfo {author} {\bibfnamefont {I.~V.}\
  \bibnamefont {Grigorieva}}, \ and\ \bibinfo {author} {\bibfnamefont {A.~A.}\
  \bibnamefont {Firsov}},\ }\bibfield  {title} {\enquote {\bibinfo {title}
  {Electric field effect in atomically thin carbon films},}\ }\href {\doibase
  10.1126/science.1102896} {\bibfield  {journal} {\bibinfo  {journal}
  {Science}\ }\textbf {\bibinfo {volume} {306}},\ \bibinfo {pages} {666--669}
  (\bibinfo {year} {2004})},\ \Eprint
  {http://arxiv.org/abs/http://www.sciencemag.org/content/306/5696/666.full.pdf}
  {http://www.sciencemag.org/content/306/5696/666.full.pdf} \BibitemShut
  {NoStop}%
\bibitem [{\citenamefont {Bunch}\ \emph {et~al.}(2007)\citenamefont {Bunch},
  \citenamefont {Van Der~Zande}, \citenamefont {Verbridge}, \citenamefont
  {Frank}, \citenamefont {Tanenbaum}, \citenamefont {Parpia}, \citenamefont
  {Craighead},\ and\ \citenamefont {McEuen}}]{Bunch2007}%
  \BibitemOpen
  \bibfield  {author} {\bibinfo {author} {\bibfnamefont {J.S.}\ \bibnamefont
  {Bunch}}, \bibinfo {author} {\bibfnamefont {A.M.}\ \bibnamefont {Van
  Der~Zande}}, \bibinfo {author} {\bibfnamefont {S.S.}\ \bibnamefont
  {Verbridge}}, \bibinfo {author} {\bibfnamefont {I.W.}\ \bibnamefont {Frank}},
  \bibinfo {author} {\bibfnamefont {D.M.}\ \bibnamefont {Tanenbaum}}, \bibinfo
  {author} {\bibfnamefont {J.M.}\ \bibnamefont {Parpia}}, \bibinfo {author}
  {\bibfnamefont {H.G.}\ \bibnamefont {Craighead}}, \ and\ \bibinfo {author}
  {\bibfnamefont {P.L.}\ \bibnamefont {McEuen}},\ }\bibfield  {title} {\enquote
  {\bibinfo {title} {Electromechanical resonators from graphene sheets},}\
  }\href {http://www.sciencemag.org/content/315/5811/490} {\bibfield  {journal}
  {\bibinfo  {journal} {Science}\ }\textbf {\bibinfo {volume} {315}},\ \bibinfo
  {pages} {490--493} (\bibinfo {year} {2007})}\BibitemShut {NoStop}%
\bibitem [{\citenamefont {Chen}\ \emph {et~al.}(2009)\citenamefont {Chen},
  \citenamefont {Rosenblatt}, \citenamefont {Bolotin}, \citenamefont {Kalb},
  \citenamefont {Kim}, \citenamefont {Kymissis}, \citenamefont {Stormer},
  \citenamefont {Heinz},\ and\ \citenamefont {Hone}}]{Chen2009}%
  \BibitemOpen
  \bibfield  {author} {\bibinfo {author} {\bibfnamefont {Changyao}\
  \bibnamefont {Chen}}, \bibinfo {author} {\bibfnamefont {Sami}\ \bibnamefont
  {Rosenblatt}}, \bibinfo {author} {\bibfnamefont {Kirill~I.}\ \bibnamefont
  {Bolotin}}, \bibinfo {author} {\bibfnamefont {William}\ \bibnamefont {Kalb}},
  \bibinfo {author} {\bibfnamefont {Philip}\ \bibnamefont {Kim}}, \bibinfo
  {author} {\bibfnamefont {Ioannis}\ \bibnamefont {Kymissis}}, \bibinfo
  {author} {\bibfnamefont {Horst~L.}\ \bibnamefont {Stormer}}, \bibinfo
  {author} {\bibfnamefont {Tony~F.}\ \bibnamefont {Heinz}}, \ and\ \bibinfo
  {author} {\bibfnamefont {James}\ \bibnamefont {Hone}},\ }\bibfield  {title}
  {\enquote {\bibinfo {title} {Performance of monolayer graphene nanomechanical
  resonators with electrical readout},}\ }\href
  {http://dx.doi.org/10.1038/nnano.2009.267} {\bibfield  {journal} {\bibinfo
  {journal} {Nat Nano}\ }\textbf {\bibinfo {volume} {4}},\ \bibinfo {pages}
  {861--867} (\bibinfo {year} {2009})}\BibitemShut {NoStop}%
\bibitem [{\citenamefont {Koppens}\ \emph {et~al.}(2011)\citenamefont
  {Koppens}, \citenamefont {Chang},\ and\ \citenamefont {Garcia~de
  Abajo}}]{Koppens2011}%
  \BibitemOpen
  \bibfield  {author} {\bibinfo {author} {\bibfnamefont {Frank H.~L.}\
  \bibnamefont {Koppens}}, \bibinfo {author} {\bibfnamefont {Darrick~E.}\
  \bibnamefont {Chang}}, \ and\ \bibinfo {author} {\bibfnamefont {F.~Javier}\
  \bibnamefont {Garcia~de Abajo}},\ }\bibfield  {title} {\enquote {\bibinfo
  {title} {Graphene plasmonics: A platform for strong light–matter
  interactions},}\ }\href {\doibase 10.1021/nl201771h} {\bibfield  {journal}
  {\bibinfo  {journal} {Nano Letters}\ }\textbf {\bibinfo {volume} {11}},\
  \bibinfo {pages} {3370--3377} (\bibinfo {year} {2011})},\ \Eprint
  {http://arxiv.org/abs/http://pubs.acs.org/doi/pdf/10.1021/nl201771h}
  {http://pubs.acs.org/doi/pdf/10.1021/nl201771h} \BibitemShut {NoStop}%
\bibitem [{\citenamefont {Nair}\ \emph {et~al.}(2008)\citenamefont {Nair},
  \citenamefont {Blake}, \citenamefont {Grigorenko}, \citenamefont {Novoselov},
  \citenamefont {Booth}, \citenamefont {Stauber}, \citenamefont {Peres},\ and\
  \citenamefont {Geim}}]{Nair2008}%
  \BibitemOpen
  \bibfield  {author} {\bibinfo {author} {\bibfnamefont {R.~R.}\ \bibnamefont
  {Nair}}, \bibinfo {author} {\bibfnamefont {P.}~\bibnamefont {Blake}},
  \bibinfo {author} {\bibfnamefont {A.~N.}\ \bibnamefont {Grigorenko}},
  \bibinfo {author} {\bibfnamefont {K.~S.}\ \bibnamefont {Novoselov}}, \bibinfo
  {author} {\bibfnamefont {T.~J.}\ \bibnamefont {Booth}}, \bibinfo {author}
  {\bibfnamefont {T.}~\bibnamefont {Stauber}}, \bibinfo {author} {\bibfnamefont
  {N.~M.~R.}\ \bibnamefont {Peres}}, \ and\ \bibinfo {author} {\bibfnamefont
  {A.~K.}\ \bibnamefont {Geim}},\ }\bibfield  {title} {\enquote {\bibinfo
  {title} {{Fine Structure Constant Defines Visual Transparency of
  Graphene}},}\ }\href {http://www.sciencemag.org/content/320/5881/1308.short}
  {\bibfield  {journal} {\bibinfo  {journal} {Science}\ }\textbf {\bibinfo
  {volume} {320}},\ \bibinfo {pages} {1308} (\bibinfo {year}
  {2008})}\BibitemShut {NoStop}%
\bibitem [{\citenamefont {Mak}\ \emph {et~al.}(2008)\citenamefont {Mak},
  \citenamefont {Sfeir}, \citenamefont {Wu}, \citenamefont {Lui}, \citenamefont
  {Misewich},\ and\ \citenamefont {Heinz}}]{Mak2008}%
  \BibitemOpen
  \bibfield  {author} {\bibinfo {author} {\bibfnamefont {Kin~Fai}\ \bibnamefont
  {Mak}}, \bibinfo {author} {\bibfnamefont {Matthew~Y.}\ \bibnamefont {Sfeir}},
  \bibinfo {author} {\bibfnamefont {Yang}\ \bibnamefont {Wu}}, \bibinfo
  {author} {\bibfnamefont {Chun~Hung}\ \bibnamefont {Lui}}, \bibinfo {author}
  {\bibfnamefont {James~A.}\ \bibnamefont {Misewich}}, \ and\ \bibinfo {author}
  {\bibfnamefont {Tony~F.}\ \bibnamefont {Heinz}},\ }\bibfield  {title}
  {\enquote {\bibinfo {title} {Measurement of the optical conductivity of
  graphene},}\ }\href
  {http://journals.aps.org/prl/abstract/10.1103/PhysRevLett.101.196405}
  {\bibfield  {journal} {\bibinfo  {journal} {Phys. Rev. Lett.}\ }\textbf
  {\bibinfo {volume} {101}},\ \bibinfo {pages} {196405} (\bibinfo {year}
  {2008})}\BibitemShut {NoStop}%
\bibitem [{\citenamefont {Castellanos-Gomez}\ \emph {et~al.}(2013)\citenamefont
  {Castellanos-Gomez}, \citenamefont {van Leeuwen}, \citenamefont {Buscema},
  \citenamefont {van~der Zant}, \citenamefont {Steele},\ and\ \citenamefont
  {Venstra}}]{Castellanos-Gomez2013}%
  \BibitemOpen
  \bibfield  {author} {\bibinfo {author} {\bibfnamefont {Andres}\ \bibnamefont
  {Castellanos-Gomez}}, \bibinfo {author} {\bibfnamefont {Ronald}\ \bibnamefont
  {van Leeuwen}}, \bibinfo {author} {\bibfnamefont {Michele}\ \bibnamefont
  {Buscema}}, \bibinfo {author} {\bibfnamefont {Herre S.~J.}\ \bibnamefont
  {van~der Zant}}, \bibinfo {author} {\bibfnamefont {Gary~A.}\ \bibnamefont
  {Steele}}, \ and\ \bibinfo {author} {\bibfnamefont {Warner~J.}\ \bibnamefont
  {Venstra}},\ }\bibfield  {title} {\enquote {\bibinfo {title} {Single-layer
  MoS$_2$ mechanical resonators},}\ }\href
  {http://dx.doi.org/10.1002/adma.201303569} {\bibfield  {journal} {\bibinfo
  {journal} {Adv. Mater.}\ }\textbf {\bibinfo {volume} {25}},\ \bibinfo {pages}
  {6719--6723} (\bibinfo {year} {2013})}\BibitemShut {NoStop}%
\bibitem [{\citenamefont {Reserbat-Plantey}\ \emph {et~al.}(2012)\citenamefont
  {Reserbat-Plantey}, \citenamefont {Marty}, \citenamefont {Arcizet},
  \citenamefont {Bendiab},\ and\ \citenamefont {Bouchiat}}]{Reserbat2012}%
  \BibitemOpen
  \bibfield  {author} {\bibinfo {author} {\bibfnamefont {Antoine}\ \bibnamefont
  {Reserbat-Plantey}}, \bibinfo {author} {\bibfnamefont {Laetitia}\
  \bibnamefont {Marty}}, \bibinfo {author} {\bibfnamefont {Olivier}\
  \bibnamefont {Arcizet}}, \bibinfo {author} {\bibfnamefont {Nedjma}\
  \bibnamefont {Bendiab}}, \ and\ \bibinfo {author} {\bibfnamefont {Vincent}\
  \bibnamefont {Bouchiat}},\ }\bibfield  {title} {\enquote {\bibinfo {title} {A
  local optical probe for measuring motion and stress in a
  nanoelectromechanical system},}\ }\href {\doibase 10.1038/nnano.2011.250}
  {\bibfield  {journal} {\bibinfo  {journal} {Nat Nano}\ }\textbf {\bibinfo
  {volume} {7}},\ \bibinfo {pages} {151--155} (\bibinfo {year}
  {2012})}\BibitemShut {NoStop}%
\bibitem [{\citenamefont {Barton}\ \emph {et~al.}(2012)\citenamefont {Barton},
  \citenamefont {Storch}, \citenamefont {Adiga}, \citenamefont {Sakakibara},
  \citenamefont {Cipriany}, \citenamefont {Ilic}, \citenamefont {Wang},
  \citenamefont {Ong}, \citenamefont {{McEuen}}, \citenamefont {Parpia},\ and\
  \citenamefont {Craighead}}]{Barton2012}%
  \BibitemOpen
  \bibfield  {author} {\bibinfo {author} {\bibfnamefont {Robert~A.}\
  \bibnamefont {Barton}}, \bibinfo {author} {\bibfnamefont {Isaac~R.}\
  \bibnamefont {Storch}}, \bibinfo {author} {\bibfnamefont {Vivekananda~P.}\
  \bibnamefont {Adiga}}, \bibinfo {author} {\bibfnamefont {Reyu}\ \bibnamefont
  {Sakakibara}}, \bibinfo {author} {\bibfnamefont {Benjamin~R.}\ \bibnamefont
  {Cipriany}}, \bibinfo {author} {\bibfnamefont {B.}~\bibnamefont {Ilic}},
  \bibinfo {author} {\bibfnamefont {Si~Ping}\ \bibnamefont {Wang}}, \bibinfo
  {author} {\bibfnamefont {Peijie}\ \bibnamefont {Ong}}, \bibinfo {author}
  {\bibfnamefont {Paul~L.}\ \bibnamefont {{McEuen}}}, \bibinfo {author}
  {\bibfnamefont {Jeevak~M.}\ \bibnamefont {Parpia}}, \ and\ \bibinfo {author}
  {\bibfnamefont {Harold~G.}\ \bibnamefont {Craighead}},\ }\bibfield  {title}
  {\enquote {\bibinfo {title} {Photothermal self-oscillation and laser cooling
  of graphene optomechanical systems},}\ }\href {\doibase 10.1021/nl302036x}
  {\bibfield  {journal} {\bibinfo  {journal} {Nano Lett.}\ }\textbf {\bibinfo
  {volume} {12}},\ \bibinfo {pages} {4681--4686} (\bibinfo {year}
  {2012})}\BibitemShut {NoStop}%
\bibitem [{\citenamefont {Koenig}\ \emph {et~al.}(2012)\citenamefont {Koenig},
  \citenamefont {Wang}, \citenamefont {Pellegrino},\ and\ \citenamefont
  {Bunch}}]{Koenig2012}%
  \BibitemOpen
  \bibfield  {author} {\bibinfo {author} {\bibfnamefont {Steven~P.}\
  \bibnamefont {Koenig}}, \bibinfo {author} {\bibfnamefont {Luda}\ \bibnamefont
  {Wang}}, \bibinfo {author} {\bibfnamefont {John}\ \bibnamefont {Pellegrino}},
  \ and\ \bibinfo {author} {\bibfnamefont {J.~Scott}\ \bibnamefont {Bunch}},\
  }\bibfield  {title} {\enquote {\bibinfo {title} {Selective molecular sieving
  through porous graphene},}\ }\href {\doibase 10.1038/nnano.2012.162}
  {\bibfield  {journal} {\bibinfo  {journal} {Nat Nano}\ }\textbf {\bibinfo
  {volume} {7}},\ \bibinfo {pages} {728--732} (\bibinfo {year}
  {2012})}\BibitemShut {NoStop}%
\bibitem [{\citenamefont {Komaragiri}\ \emph {et~al.}(2005)\citenamefont
  {Komaragiri}, \citenamefont {Begley},\ and\ \citenamefont
  {Simmonds}}]{Komaragiri2005}%
  \BibitemOpen
  \bibfield  {author} {\bibinfo {author} {\bibfnamefont {U.}~\bibnamefont
  {Komaragiri}}, \bibinfo {author} {\bibfnamefont {M.~R.}\ \bibnamefont
  {Begley}}, \ and\ \bibinfo {author} {\bibfnamefont {J.~G.}\ \bibnamefont
  {Simmonds}},\ }\bibfield  {title} {\enquote {\bibinfo {title} {The mechanical
  response of freestanding circular elastic films under point and pressure
  loads},}\ }\href {\doibase 10.1115/1.1827246} {\bibfield  {journal} {\bibinfo
   {journal} {J. Appl. Mech.}\ }\textbf {\bibinfo {volume} {72}},\ \bibinfo
  {pages} {203--212} (\bibinfo {year} {2005})}\BibitemShut {NoStop}%
\bibitem [{\citenamefont {Georgiou}\ \emph {et~al.}(2011)\citenamefont
  {Georgiou}, \citenamefont {Britnell}, \citenamefont {Blake}, \citenamefont
  {Gorbachev}, \citenamefont {Gholinia}, \citenamefont {Geim}, \citenamefont
  {Casiraghi},\ and\ \citenamefont {Novoselov}}]{Georgiou2011}%
  \BibitemOpen
  \bibfield  {author} {\bibinfo {author} {\bibfnamefont {T.}~\bibnamefont
  {Georgiou}}, \bibinfo {author} {\bibfnamefont {L.}~\bibnamefont {Britnell}},
  \bibinfo {author} {\bibfnamefont {P.}~\bibnamefont {Blake}}, \bibinfo
  {author} {\bibfnamefont {R.~V.}\ \bibnamefont {Gorbachev}}, \bibinfo {author}
  {\bibfnamefont {A.}~\bibnamefont {Gholinia}}, \bibinfo {author}
  {\bibfnamefont {A.~K.}\ \bibnamefont {Geim}}, \bibinfo {author}
  {\bibfnamefont {C.}~\bibnamefont {Casiraghi}}, \ and\ \bibinfo {author}
  {\bibfnamefont {K.~S.}\ \bibnamefont {Novoselov}},\ }\bibfield  {title}
  {\enquote {\bibinfo {title} {Graphene bubbles with controllable curvature},}\
  }\href {\doibase http://dx.doi.org/10.1063/1.3631632} {\bibfield  {journal}
  {\bibinfo  {journal} {Applied Physics Letters}\ }\textbf {\bibinfo {volume}
  {99}},\ \bibinfo {pages} {093103} (\bibinfo {year} {2011})}\BibitemShut
  {NoStop}%
\bibitem [{\citenamefont {Zabel}\ \emph {et~al.}(2011)\citenamefont {Zabel},
  \citenamefont {Nair}, \citenamefont {Ott}, \citenamefont {Georgiou},
  \citenamefont {Geim}, \citenamefont {Novoselov},\ and\ \citenamefont
  {Casiraghi}}]{Zabel2011}%
  \BibitemOpen
  \bibfield  {author} {\bibinfo {author} {\bibfnamefont {Jakob}\ \bibnamefont
  {Zabel}}, \bibinfo {author} {\bibfnamefont {Rahul~R.}\ \bibnamefont {Nair}},
  \bibinfo {author} {\bibfnamefont {Anna}\ \bibnamefont {Ott}}, \bibinfo
  {author} {\bibfnamefont {Thanasis}\ \bibnamefont {Georgiou}}, \bibinfo
  {author} {\bibfnamefont {Andre~K.}\ \bibnamefont {Geim}}, \bibinfo {author}
  {\bibfnamefont {Kostya~S.}\ \bibnamefont {Novoselov}}, \ and\ \bibinfo
  {author} {\bibfnamefont {Cinzia}\ \bibnamefont {Casiraghi}},\ }\bibfield
  {title} {\enquote {\bibinfo {title} {Raman spectroscopy of graphene and
  bilayer under biaxial strain: Bubbles and balloons},}\ }\href {\doibase
  10.1021/nl203359n} {\bibfield  {journal} {\bibinfo  {journal} {Nano Lett.}\
  }\textbf {\bibinfo {volume} {12}},\ \bibinfo {pages} {617--621} (\bibinfo
  {year} {2011})}\BibitemShut {NoStop}%
\bibitem [{\citenamefont {Lee}\ \emph {et~al.}(2012{\natexlab{a}})\citenamefont
  {Lee}, \citenamefont {Yoon},\ and\ \citenamefont {Cheong}}]{Lee2012}%
  \BibitemOpen
  \bibfield  {author} {\bibinfo {author} {\bibfnamefont {Jae-Ung}\ \bibnamefont
  {Lee}}, \bibinfo {author} {\bibfnamefont {Duhee}\ \bibnamefont {Yoon}}, \
  and\ \bibinfo {author} {\bibfnamefont {Hyeonsik}\ \bibnamefont {Cheong}},\
  }\bibfield  {title} {\enquote {\bibinfo {title} {Estimation of young's
  modulus of graphene by raman spectroscopy},}\ }\href {\doibase
  10.1021/nl301073q} {\bibfield  {journal} {\bibinfo  {journal} {Nano Letters}\
  }\textbf {\bibinfo {volume} {12}},\ \bibinfo {pages} {4444--4448} (\bibinfo
  {year} {2012}{\natexlab{a}})}\BibitemShut {NoStop}%
\bibitem [{\citenamefont {Kitt}\ \emph {et~al.}(2013)\citenamefont {Kitt},
  \citenamefont {Qi}, \citenamefont {R?©mi}, \citenamefont {Park},
  \citenamefont {Swan},\ and\ \citenamefont {Goldberg}}]{Kitt2013}%
  \BibitemOpen
  \bibfield  {author} {\bibinfo {author} {\bibfnamefont {Alexander~L.}\
  \bibnamefont {Kitt}}, \bibinfo {author} {\bibfnamefont {Zenan}\ \bibnamefont
  {Qi}}, \bibinfo {author} {\bibfnamefont {Sebastian}\ \bibnamefont {R?©mi}},
  \bibinfo {author} {\bibfnamefont {Harold~S.}\ \bibnamefont {Park}}, \bibinfo
  {author} {\bibfnamefont {Anna~K.}\ \bibnamefont {Swan}}, \ and\ \bibinfo
  {author} {\bibfnamefont {Bennett~B.}\ \bibnamefont {Goldberg}},\ }\bibfield
  {title} {\enquote {\bibinfo {title} {How graphene slides: Measurement and
  theory of strain-dependent frictional forces between graphene and SiO$_2$},}\
  }\href {\doibase 10.1021/nl4007112} {\bibfield  {journal} {\bibinfo
  {journal} {Nano Letters}\ }\textbf {\bibinfo {volume} {13}},\ \bibinfo
  {pages} {2605--2610} (\bibinfo {year} {2013})}\BibitemShut {NoStop}%
\bibitem [{\citenamefont {Berciaud}\ \emph {et~al.}(2009)\citenamefont
  {Berciaud}, \citenamefont {Ryu}, \citenamefont {Brus},\ and\ \citenamefont
  {Heinz}}]{Berciaud2009}%
  \BibitemOpen
  \bibfield  {author} {\bibinfo {author} {\bibfnamefont {S.}~\bibnamefont
  {Berciaud}}, \bibinfo {author} {\bibfnamefont {S.}~\bibnamefont {Ryu}},
  \bibinfo {author} {\bibfnamefont {L.~E.}\ \bibnamefont {Brus}}, \ and\
  \bibinfo {author} {\bibfnamefont {T.~F.}\ \bibnamefont {Heinz}},\ }\bibfield
  {title} {\enquote {\bibinfo {title} {Probing the intrinsic properties of
  exfoliated graphene: Raman spectroscopy of free-standing monolayers},}\
  }\href {\doibase 10.1021/nl8031444} {\bibfield  {journal} {\bibinfo
  {journal} {Nano Letters}\ }\textbf {\bibinfo {volume} {9}},\ \bibinfo {pages}
  {346--352} (\bibinfo {year} {2009})}\BibitemShut {NoStop}%
\bibitem [{SMn()}]{SMnote}%
  \BibitemOpen
  \href@noop {} {}\bibinfo {note} {See Supplemental Material for details on the
  determination of the pressure load, the determination of the blister height
  from the intensity of the 2D mode feature, the influence of the laser
  focusing conditions, polarization resolved Raman measurements, data measured
  on other samples and measurements on a concave graphene blister.}\BibitemShut
  {Stop}%
\bibitem [{\citenamefont {Malard}\ \emph {et~al.}(2009)\citenamefont {Malard},
  \citenamefont {Pimenta}, \citenamefont {Dresselhaus},\ and\ \citenamefont
  {Dresselhaus}}]{Malard2009}%
  \BibitemOpen
  \bibfield  {author} {\bibinfo {author} {\bibfnamefont {L.M.}\ \bibnamefont
  {Malard}}, \bibinfo {author} {\bibfnamefont {M.A.}\ \bibnamefont {Pimenta}},
  \bibinfo {author} {\bibfnamefont {G.}~\bibnamefont {Dresselhaus}}, \ and\
  \bibinfo {author} {\bibfnamefont {M.S.}\ \bibnamefont {Dresselhaus}},\
  }\bibfield  {title} {\enquote {\bibinfo {title} {Raman spectroscopy in
  graphene},}\ }\href
  {http://www.sciencedirect.com/science/article/pii/S0370157309000520}
  {\bibfield  {journal} {\bibinfo  {journal} {Physics Reports}\ }\textbf
  {\bibinfo {volume} {473}},\ \bibinfo {pages} {51--87} (\bibinfo {year}
  {2009})}\BibitemShut {NoStop}%
\bibitem [{\citenamefont {Ferrari}\ and\ \citenamefont
  {Basko}(2013)}]{Ferrari2013}%
  \BibitemOpen
  \bibfield  {author} {\bibinfo {author} {\bibfnamefont {Andrea~C.}\
  \bibnamefont {Ferrari}}\ and\ \bibinfo {author} {\bibfnamefont {Denis~M.}\
  \bibnamefont {Basko}},\ }\bibfield  {title} {\enquote {\bibinfo {title}
  {Raman spectroscopy as a versatile tool for studying the properties of
  graphene},}\ }\href {http://dx.doi.org/10.1038/nnano.2013.46} {\bibfield
  {journal} {\bibinfo  {journal} {Nat Nano}\ }\textbf {\bibinfo {volume} {8}},\
  \bibinfo {pages} {235--246} (\bibinfo {year} {2013})}\BibitemShut {NoStop}%
\bibitem [{\citenamefont {Cancado}\ \emph {et~al.}(2011)\citenamefont
  {Cancado}, \citenamefont {Jorio}, \citenamefont {Ferreira}, \citenamefont
  {Stavale}, \citenamefont {Achete}, \citenamefont {Capaz}, \citenamefont
  {Moutinho}, \citenamefont {Lombardo}, \citenamefont {Kulmala},\ and\
  \citenamefont {Ferrari}}]{Cancado2011}%
  \BibitemOpen
  \bibfield  {author} {\bibinfo {author} {\bibfnamefont {L.~G.}\ \bibnamefont
  {Cancado}}, \bibinfo {author} {\bibfnamefont {A.}~\bibnamefont {Jorio}},
  \bibinfo {author} {\bibfnamefont {E.~H.~Martins}\ \bibnamefont {Ferreira}},
  \bibinfo {author} {\bibfnamefont {F.}~\bibnamefont {Stavale}}, \bibinfo
  {author} {\bibfnamefont {C.~A.}\ \bibnamefont {Achete}}, \bibinfo {author}
  {\bibfnamefont {R.~B.}\ \bibnamefont {Capaz}}, \bibinfo {author}
  {\bibfnamefont {M.~V.~O.}\ \bibnamefont {Moutinho}}, \bibinfo {author}
  {\bibfnamefont {A.}~\bibnamefont {Lombardo}}, \bibinfo {author}
  {\bibfnamefont {T.~S.}\ \bibnamefont {Kulmala}}, \ and\ \bibinfo {author}
  {\bibfnamefont {A.~C.}\ \bibnamefont {Ferrari}},\ }\bibfield  {title}
  {\enquote {\bibinfo {title} {Quantifying defects in graphene via raman
  spectroscopy at different excitation energies},}\ }\href {\doibase
  10.1021/nl201432g} {\bibfield  {journal} {\bibinfo  {journal} {Nano Lett.}\
  }\textbf {\bibinfo {volume} {11}},\ \bibinfo {pages} {3190--3196} (\bibinfo
  {year} {2011})}\BibitemShut {NoStop}%
\bibitem [{\citenamefont {Eckmann}\ \emph {et~al.}(2012)\citenamefont
  {Eckmann}, \citenamefont {Felten}, \citenamefont {Mishchenko}, \citenamefont
  {Britnell}, \citenamefont {Krupke}, \citenamefont {Novoselov},\ and\
  \citenamefont {Casiraghi}}]{Eckmann2012}%
  \BibitemOpen
  \bibfield  {author} {\bibinfo {author} {\bibfnamefont {Axel}\ \bibnamefont
  {Eckmann}}, \bibinfo {author} {\bibfnamefont {Alexandre}\ \bibnamefont
  {Felten}}, \bibinfo {author} {\bibfnamefont {Artem}\ \bibnamefont
  {Mishchenko}}, \bibinfo {author} {\bibfnamefont {Liam}\ \bibnamefont
  {Britnell}}, \bibinfo {author} {\bibfnamefont {Ralph}\ \bibnamefont
  {Krupke}}, \bibinfo {author} {\bibfnamefont {Kostya~S.}\ \bibnamefont
  {Novoselov}}, \ and\ \bibinfo {author} {\bibfnamefont {Cinzia}\ \bibnamefont
  {Casiraghi}},\ }\bibfield  {title} {\enquote {\bibinfo {title} {Probing the
  nature of defects in graphene by raman spectroscopy},}\ }\href {\doibase
  10.1021/nl300901a} {\bibfield  {journal} {\bibinfo  {journal} {Nano Lett.}\
  }\textbf {\bibinfo {volume} {12}},\ \bibinfo {pages} {3925--3930} (\bibinfo
  {year} {2012})}\BibitemShut {NoStop}%
\bibitem [{\citenamefont {Mohiuddin}\ \emph {et~al.}(2009)\citenamefont
  {Mohiuddin}, \citenamefont {Lombardo}, \citenamefont {Nair}, \citenamefont
  {Bonetti}, \citenamefont {Savini}, \citenamefont {Jalil}, \citenamefont
  {Bonini}, \citenamefont {Basko}, \citenamefont {Galiotis}, \citenamefont
  {Marzari}, \citenamefont {Novoselov}, \citenamefont {Geim},\ and\
  \citenamefont {Ferrari}}]{Mohiuddin2009}%
  \BibitemOpen
  \bibfield  {author} {\bibinfo {author} {\bibfnamefont {T.~M.~G.}\
  \bibnamefont {Mohiuddin}}, \bibinfo {author} {\bibfnamefont {A.}~\bibnamefont
  {Lombardo}}, \bibinfo {author} {\bibfnamefont {R.~R.}\ \bibnamefont {Nair}},
  \bibinfo {author} {\bibfnamefont {A.}~\bibnamefont {Bonetti}}, \bibinfo
  {author} {\bibfnamefont {G.}~\bibnamefont {Savini}}, \bibinfo {author}
  {\bibfnamefont {R.}~\bibnamefont {Jalil}}, \bibinfo {author} {\bibfnamefont
  {N.}~\bibnamefont {Bonini}}, \bibinfo {author} {\bibfnamefont {D.~M.}\
  \bibnamefont {Basko}}, \bibinfo {author} {\bibfnamefont {C.}~\bibnamefont
  {Galiotis}}, \bibinfo {author} {\bibfnamefont {N.}~\bibnamefont {Marzari}},
  \bibinfo {author} {\bibfnamefont {K.~S.}\ \bibnamefont {Novoselov}}, \bibinfo
  {author} {\bibfnamefont {A.~K.}\ \bibnamefont {Geim}}, \ and\ \bibinfo
  {author} {\bibfnamefont {A.~C.}\ \bibnamefont {Ferrari}},\ }\bibfield
  {title} {\enquote {\bibinfo {title} {Uniaxial strain in graphene by raman
  spectroscopy: $G$ peak splitting, gr\"uneisen parameters, and sample
  orientation},}\ }\href {\doibase 10.1103/PhysRevB.79.205433} {\bibfield
  {journal} {\bibinfo  {journal} {Phys. Rev. B}\ }\textbf {\bibinfo {volume}
  {79}},\ \bibinfo {pages} {205433} (\bibinfo {year} {2009})}\BibitemShut
  {NoStop}%
\bibitem [{\citenamefont {Huang}\ \emph {et~al.}(2009)\citenamefont {Huang},
  \citenamefont {Yan}, \citenamefont {Chen}, \citenamefont {Song},
  \citenamefont {Heinz},\ and\ \citenamefont {Hone}}]{Huang2009}%
  \BibitemOpen
  \bibfield  {author} {\bibinfo {author} {\bibfnamefont {M.}~\bibnamefont
  {Huang}}, \bibinfo {author} {\bibfnamefont {H.}~\bibnamefont {Yan}}, \bibinfo
  {author} {\bibfnamefont {C.}~\bibnamefont {Chen}}, \bibinfo {author}
  {\bibfnamefont {D.}~\bibnamefont {Song}}, \bibinfo {author} {\bibfnamefont
  {Tony~F.}\ \bibnamefont {Heinz}}, \ and\ \bibinfo {author} {\bibfnamefont
  {J.}~\bibnamefont {Hone}},\ }\bibfield  {title} {\enquote {\bibinfo {title}
  {Phonon softening and crystallographic orientation of strained graphene
  studied by raman spectroscopy},}\ }\href
  {http://www.pnas.org/content/106/18/7304.full} {\bibfield  {journal}
  {\bibinfo  {journal} {Proc. Natl. Acad. Sci. U.S.A.}\ }\textbf {\bibinfo
  {volume} {106}},\ \bibinfo {pages} {7304} (\bibinfo {year}
  {2009})}\BibitemShut {NoStop}%
\bibitem [{\citenamefont {Metzger}\ \emph {et~al.}(2009)\citenamefont
  {Metzger}, \citenamefont {R\'emi}, \citenamefont {Liu}, \citenamefont
  {Kusminskiy}, \citenamefont {Castro~Neto}, \citenamefont {Swan},\ and\
  \citenamefont {Goldberg}}]{Metzger2009}%
  \BibitemOpen
  \bibfield  {author} {\bibinfo {author} {\bibfnamefont {Constanze}\
  \bibnamefont {Metzger}}, \bibinfo {author} {\bibfnamefont {Sebastian}\
  \bibnamefont {R\'emi}}, \bibinfo {author} {\bibfnamefont {Mengkun}\
  \bibnamefont {Liu}}, \bibinfo {author} {\bibfnamefont {Silvia~V.}\
  \bibnamefont {Kusminskiy}}, \bibinfo {author} {\bibfnamefont {Antonio~H.}\
  \bibnamefont {Castro~Neto}}, \bibinfo {author} {\bibfnamefont {Anna~K.}\
  \bibnamefont {Swan}}, \ and\ \bibinfo {author} {\bibfnamefont {Bennett~B.}\
  \bibnamefont {Goldberg}},\ }\bibfield  {title} {\enquote {\bibinfo {title}
  {Biaxial strain in graphene adhered to shallow depressions},}\ }\href
  {\doibase 10.1021/nl901625v} {\bibfield  {journal} {\bibinfo  {journal} {Nano
  Lett.}\ }\textbf {\bibinfo {volume} {10}},\ \bibinfo {pages} {6--10}
  (\bibinfo {year} {2009})}\BibitemShut {NoStop}%
\bibitem [{\citenamefont {Ding}\ \emph {et~al.}(2010)\citenamefont {Ding},
  \citenamefont {Ji}, \citenamefont {Chen}, \citenamefont {Herklotz},
  \citenamefont {D\"{o}rr}, \citenamefont {Mei}, \citenamefont {Rastelli},\
  and\ \citenamefont {Schmidt}}]{Ding2010}%
  \BibitemOpen
  \bibfield  {author} {\bibinfo {author} {\bibfnamefont {Fei}\ \bibnamefont
  {Ding}}, \bibinfo {author} {\bibfnamefont {Hengxing}\ \bibnamefont {Ji}},
  \bibinfo {author} {\bibfnamefont {Yonghai}\ \bibnamefont {Chen}}, \bibinfo
  {author} {\bibfnamefont {Andreas}\ \bibnamefont {Herklotz}}, \bibinfo
  {author} {\bibfnamefont {Kathrin}\ \bibnamefont {D\"{o}rr}}, \bibinfo
  {author} {\bibfnamefont {Yongfeng}\ \bibnamefont {Mei}}, \bibinfo {author}
  {\bibfnamefont {Armando}\ \bibnamefont {Rastelli}}, \ and\ \bibinfo {author}
  {\bibfnamefont {Oliver~G.}\ \bibnamefont {Schmidt}},\ }\bibfield  {title}
  {\enquote {\bibinfo {title} {Stretchable graphene: A close look at
  fundamental parameters through biaxial straining},}\ }\href {\doibase
  10.1021/nl101533x} {\bibfield  {journal} {\bibinfo  {journal} {Nano Lett.}\
  }\textbf {\bibinfo {volume} {10}},\ \bibinfo {pages} {3453--3458} (\bibinfo
  {year} {2010})}\BibitemShut {NoStop}%
\bibitem [{\citenamefont {Lee}\ \emph {et~al.}(2012{\natexlab{b}})\citenamefont
  {Lee}, \citenamefont {Ahn}, \citenamefont {Shim}, \citenamefont {Lee},\ and\
  \citenamefont {Ryu}}]{Lee2012a}%
  \BibitemOpen
  \bibfield  {author} {\bibinfo {author} {\bibfnamefont {Ji~Eun}\ \bibnamefont
  {Lee}}, \bibinfo {author} {\bibfnamefont {Gwanghyun}\ \bibnamefont {Ahn}},
  \bibinfo {author} {\bibfnamefont {Jihye}\ \bibnamefont {Shim}}, \bibinfo
  {author} {\bibfnamefont {Young~Sik}\ \bibnamefont {Lee}}, \ and\ \bibinfo
  {author} {\bibfnamefont {Sunmin}\ \bibnamefont {Ryu}},\ }\bibfield  {title}
  {\enquote {\bibinfo {title} {Optical separation of mechanical strain from
  charge doping in graphene},}\ }\href {http://dx.doi.org/10.1038/ncomms2022}
  {\bibfield  {journal} {\bibinfo  {journal} {Nat Commun}\ }\textbf {\bibinfo
  {volume} {3}},\ \bibinfo {pages} {1024} (\bibinfo {year}
  {2012}{\natexlab{b}})}\BibitemShut {NoStop}%
\bibitem [{\citenamefont {Huang}\ \emph {et~al.}(2010)\citenamefont {Huang},
  \citenamefont {Yan}, \citenamefont {Heinz},\ and\ \citenamefont
  {Hone}}]{Huang2010}%
  \BibitemOpen
  \bibfield  {author} {\bibinfo {author} {\bibfnamefont {M.}~\bibnamefont
  {Huang}}, \bibinfo {author} {\bibfnamefont {H.}~\bibnamefont {Yan}}, \bibinfo
  {author} {\bibfnamefont {T.F.}\ \bibnamefont {Heinz}}, \ and\ \bibinfo
  {author} {\bibfnamefont {J.}~\bibnamefont {Hone}},\ }\bibfield  {title}
  {\enquote {\bibinfo {title} {Probing strain-induced electronic structure
  change in graphene by raman spectroscopy},}\ }\href
  {http://pubs.acs.org/doi/abs/10.1021/nl102123c} {\bibfield  {journal}
  {\bibinfo  {journal} {Nano Letters}\ }\textbf {\bibinfo {volume} {10}},\
  \bibinfo {pages} {4074} (\bibinfo {year} {2010})}\BibitemShut {NoStop}%
\bibitem [{\citenamefont {Yoon}\ \emph {et~al.}(2011)\citenamefont {Yoon},
  \citenamefont {Son},\ and\ \citenamefont {Cheong}}]{Yoon2011}%
  \BibitemOpen
  \bibfield  {author} {\bibinfo {author} {\bibfnamefont {Duhee}\ \bibnamefont
  {Yoon}}, \bibinfo {author} {\bibfnamefont {Young-Woo}\ \bibnamefont {Son}}, \
  and\ \bibinfo {author} {\bibfnamefont {Hyeonsik}\ \bibnamefont {Cheong}},\
  }\bibfield  {title} {\enquote {\bibinfo {title} {Strain-dependent splitting
  of the double-resonance raman scattering band in graphene},}\ }\href
  {http://link.aps.org/doi/10.1103/PhysRevLett.106.155502} {\bibfield
  {journal} {\bibinfo  {journal} {Phys. Rev. Lett.}\ }\textbf {\bibinfo
  {volume} {106}},\ \bibinfo {pages} {155502--} (\bibinfo {year}
  {2011})}\BibitemShut {NoStop}%
\bibitem [{\citenamefont {Frank}\ \emph {et~al.}(2011)\citenamefont {Frank},
  \citenamefont {Mohr}, \citenamefont {Maultzsch}, \citenamefont {Thomsen},
  \citenamefont {Riaz}, \citenamefont {Jalil}, \citenamefont {Novoselov},
  \citenamefont {Tsoukleri}, \citenamefont {Parthenios}, \citenamefont
  {Papagelis}, \citenamefont {Kavan},\ and\ \citenamefont
  {Galiotis}}]{Frank2011}%
  \BibitemOpen
  \bibfield  {author} {\bibinfo {author} {\bibfnamefont {Otakar}\ \bibnamefont
  {Frank}}, \bibinfo {author} {\bibfnamefont {Marcel}\ \bibnamefont {Mohr}},
  \bibinfo {author} {\bibfnamefont {Janina}\ \bibnamefont {Maultzsch}},
  \bibinfo {author} {\bibfnamefont {Christian}\ \bibnamefont {Thomsen}},
  \bibinfo {author} {\bibfnamefont {Ibtsam}\ \bibnamefont {Riaz}}, \bibinfo
  {author} {\bibfnamefont {Rashid}\ \bibnamefont {Jalil}}, \bibinfo {author}
  {\bibfnamefont {Kostya~S.}\ \bibnamefont {Novoselov}}, \bibinfo {author}
  {\bibfnamefont {Georgia}\ \bibnamefont {Tsoukleri}}, \bibinfo {author}
  {\bibfnamefont {John}\ \bibnamefont {Parthenios}}, \bibinfo {author}
  {\bibfnamefont {Konstantinos}\ \bibnamefont {Papagelis}}, \bibinfo {author}
  {\bibfnamefont {Ladislav}\ \bibnamefont {Kavan}}, \ and\ \bibinfo {author}
  {\bibfnamefont {Costas}\ \bibnamefont {Galiotis}},\ }\bibfield  {title}
  {\enquote {\bibinfo {title} {Raman 2D-band splitting in graphene: Theory and
  experiment},}\ }\href {\doibase 10.1021/nn103493g} {\bibfield  {journal}
  {\bibinfo  {journal} {ACS Nano}\ }\textbf {\bibinfo {volume} {5}},\ \bibinfo
  {pages} {2231--2239} (\bibinfo {year} {2011})}\BibitemShut {NoStop}%
\bibitem [{\citenamefont {Berciaud}\ \emph {et~al.}(2013)\citenamefont
  {Berciaud}, \citenamefont {Li}, \citenamefont {Htoon}, \citenamefont {Brus},
  \citenamefont {Doorn},\ and\ \citenamefont {Heinz}}]{Berciaud2013}%
  \BibitemOpen
  \bibfield  {author} {\bibinfo {author} {\bibfnamefont {S.}~\bibnamefont
  {Berciaud}}, \bibinfo {author} {\bibfnamefont {X.}~\bibnamefont {Li}},
  \bibinfo {author} {\bibfnamefont {H.}~\bibnamefont {Htoon}}, \bibinfo
  {author} {\bibfnamefont {L.E.}\ \bibnamefont {Brus}}, \bibinfo {author}
  {\bibfnamefont {S.K.}\ \bibnamefont {Doorn}}, \ and\ \bibinfo {author}
  {\bibfnamefont {T.F.}\ \bibnamefont {Heinz}},\ }\bibfield  {title} {\enquote
  {\bibinfo {title} {Intrinsic line shape of the raman 2D-mode in freestanding
  graphene monolayers},}\ }\href {\doibase 10.1021/nl400917e} {\bibfield
  {journal} {\bibinfo  {journal} {Nano Letters}\ }\textbf {\bibinfo {volume}
  {13}},\ \bibinfo {pages} {3517--3523} (\bibinfo {year} {2013})}\BibitemShut
  {NoStop}%
\bibitem [{\citenamefont {Ryu}\ \emph {et~al.}(2010)\citenamefont {Ryu},
  \citenamefont {Liu}, \citenamefont {Berciaud}, \citenamefont {Yu},
  \citenamefont {Liu}, \citenamefont {Kim}, \citenamefont {Flynn},\ and\
  \citenamefont {Brus}}]{Ryu2010}%
  \BibitemOpen
  \bibfield  {author} {\bibinfo {author} {\bibfnamefont {Sunmin}\ \bibnamefont
  {Ryu}}, \bibinfo {author} {\bibfnamefont {Li}~\bibnamefont {Liu}}, \bibinfo
  {author} {\bibfnamefont {Stephane}\ \bibnamefont {Berciaud}}, \bibinfo
  {author} {\bibfnamefont {Young-Jun}\ \bibnamefont {Yu}}, \bibinfo {author}
  {\bibfnamefont {Haitao}\ \bibnamefont {Liu}}, \bibinfo {author}
  {\bibfnamefont {Philip}\ \bibnamefont {Kim}}, \bibinfo {author}
  {\bibfnamefont {George~W.}\ \bibnamefont {Flynn}}, \ and\ \bibinfo {author}
  {\bibfnamefont {Louis~E.}\ \bibnamefont {Brus}},\ }\bibfield  {title}
  {\enquote {\bibinfo {title} {Atmospheric oxygen binding and hole doping in
  deformed graphene on a SiO$_2$ substrate},}\ }\href {\doibase 10.1021/nl1029607}
  {\bibfield  {journal} {\bibinfo  {journal} {Nano Letters}\ }\textbf {\bibinfo
  {volume} {10}},\ \bibinfo {pages} {4944--4951} (\bibinfo {year}
  {2010})}\BibitemShut {NoStop}%
\bibitem [{\citenamefont {Metten}\ \emph {et~al.}(2013)\citenamefont {Metten},
  \citenamefont {Federspiel}, \citenamefont {Romeo},\ and\ \citenamefont
  {Berciaud}}]{Metten2013}%
  \BibitemOpen
  \bibfield  {author} {\bibinfo {author} {\bibfnamefont {D.}~\bibnamefont
  {Metten}}, \bibinfo {author} {\bibfnamefont {F.}~\bibnamefont {Federspiel}},
  \bibinfo {author} {\bibfnamefont {M.}~\bibnamefont {Romeo}}, \ and\ \bibinfo
  {author} {\bibfnamefont {S.}~\bibnamefont {Berciaud}},\ }\bibfield  {title}
  {\enquote {\bibinfo {title} {Probing built-in strain in freestanding graphene
  monolayers by raman spectroscopy},}\ }\href
  {http://onlinelibrary.wiley.com/doi/10.1002/pssb.201300220/abstract}
  {\bibfield  {journal} {\bibinfo  {journal} {physica status solidi B}\
  }\textbf {\bibinfo {volume} {250}},\ \bibinfo {pages} {2681--2688} (\bibinfo
  {year} {2013})}\BibitemShut {NoStop}%
\bibitem [{\citenamefont {Calizo}\ \emph {et~al.}(2007)\citenamefont {Calizo},
  \citenamefont {Balandin}, \citenamefont {Bao}, \citenamefont {Miao},\ and\
  \citenamefont {Lau}}]{Calizo2007}%
  \BibitemOpen
  \bibfield  {author} {\bibinfo {author} {\bibfnamefont {I.}~\bibnamefont
  {Calizo}}, \bibinfo {author} {\bibfnamefont {A.~A.}\ \bibnamefont
  {Balandin}}, \bibinfo {author} {\bibfnamefont {W.}~\bibnamefont {Bao}},
  \bibinfo {author} {\bibfnamefont {F.}~\bibnamefont {Miao}}, \ and\ \bibinfo
  {author} {\bibfnamefont {C.~N.}\ \bibnamefont {Lau}},\ }\bibfield  {title}
  {\enquote {\bibinfo {title} {Temperature dependence of the raman spectra of
  graphene and graphene multilayers},}\ }\href {\doibase 10.1021/nl071033g}
  {\bibfield  {journal} {\bibinfo  {journal} {Nano Letters}\ }\textbf {\bibinfo
  {volume} {7}},\ \bibinfo {pages} {2645--2649} (\bibinfo {year} {2007})},\
  \Eprint {http://arxiv.org/abs/http://pubs.acs.org/doi/pdf/10.1021/nl071033g}
  {http://pubs.acs.org/doi/pdf/10.1021/nl071033g} \BibitemShut {NoStop}%
\bibitem [{\citenamefont {Blake}\ \emph {et~al.}(2007)\citenamefont {Blake},
  \citenamefont {Hill}, \citenamefont {Castro~Neto}, \citenamefont {Novoselov},
  \citenamefont {Jiang}, \citenamefont {Yang}, \citenamefont {Booth},\ and\
  \citenamefont {Geim}}]{Blake2007}%
  \BibitemOpen
  \bibfield  {author} {\bibinfo {author} {\bibfnamefont {P.}~\bibnamefont
  {Blake}}, \bibinfo {author} {\bibfnamefont {E.~W.}\ \bibnamefont {Hill}},
  \bibinfo {author} {\bibfnamefont {A.~H.}\ \bibnamefont {Castro~Neto}},
  \bibinfo {author} {\bibfnamefont {K.~S.}\ \bibnamefont {Novoselov}}, \bibinfo
  {author} {\bibfnamefont {D.}~\bibnamefont {Jiang}}, \bibinfo {author}
  {\bibfnamefont {R.}~\bibnamefont {Yang}}, \bibinfo {author} {\bibfnamefont
  {T.~J.}\ \bibnamefont {Booth}}, \ and\ \bibinfo {author} {\bibfnamefont
  {A.~K.}\ \bibnamefont {Geim}},\ }\bibfield  {title} {\enquote {\bibinfo
  {title} {Making graphene visible},}\ }\href {\doibase
  http://dx.doi.org/10.1063/1.2768624} {\bibfield  {journal} {\bibinfo
  {journal} {Applied Physics Letters}\ }\textbf {\bibinfo {volume} {91}},\
  \bibinfo {pages} {063124} (\bibinfo {year} {2007})}\BibitemShut {NoStop}%
\bibitem [{\citenamefont {Yoon}\ \emph {et~al.}(2009)\citenamefont {Yoon},
  \citenamefont {Moon}, \citenamefont {Son}, \citenamefont {Choi},
  \citenamefont {Park}, \citenamefont {Cha}, \citenamefont {Kim},\ and\
  \citenamefont {Cheong}}]{Yoon2009}%
  \BibitemOpen
  \bibfield  {author} {\bibinfo {author} {\bibfnamefont {Duhee}\ \bibnamefont
  {Yoon}}, \bibinfo {author} {\bibfnamefont {Hyerim}\ \bibnamefont {Moon}},
  \bibinfo {author} {\bibfnamefont {Young-Woo}\ \bibnamefont {Son}}, \bibinfo
  {author} {\bibfnamefont {Jin~Sik}\ \bibnamefont {Choi}}, \bibinfo {author}
  {\bibfnamefont {Bae~Ho}\ \bibnamefont {Park}}, \bibinfo {author}
  {\bibfnamefont {Young~Hun}\ \bibnamefont {Cha}}, \bibinfo {author}
  {\bibfnamefont {Young~Dong}\ \bibnamefont {Kim}}, \ and\ \bibinfo {author}
  {\bibfnamefont {Hyeonsik}\ \bibnamefont {Cheong}},\ }\bibfield  {title}
  {\enquote {\bibinfo {title} {Interference effect on raman spectrum of
  graphene on SiO$2$/Si},}\ }\href
  {http://link.aps.org/doi/10.1103/PhysRevB.80.125422} {\bibfield  {journal}
  {\bibinfo  {journal} {Phys. Rev. B}\ }\textbf {\bibinfo {volume} {80}},\
  \bibinfo {pages} {125422--} (\bibinfo {year} {2009})}\BibitemShut {NoStop}%
\bibitem [{\citenamefont {Reserbat-Plantey}\ \emph {et~al.}(2013)\citenamefont
  {Reserbat-Plantey}, \citenamefont {Klyatskaya}, \citenamefont {Reita},
  \citenamefont {Marty}, \citenamefont {Arcizet}, \citenamefont {Ruben},
  \citenamefont {Bendiab},\ and\ \citenamefont
  {Bouchiat}}]{Reserbat-Plantey2013}%
  \BibitemOpen
  \bibfield  {author} {\bibinfo {author} {\bibfnamefont {Antoine}\ \bibnamefont
  {Reserbat-Plantey}}, \bibinfo {author} {\bibfnamefont {Svetlana}\
  \bibnamefont {Klyatskaya}}, \bibinfo {author} {\bibfnamefont {Val\'erie}\
  \bibnamefont {Reita}}, \bibinfo {author} {\bibfnamefont {Laetitia}\
  \bibnamefont {Marty}}, \bibinfo {author} {\bibfnamefont {Olivier}\
  \bibnamefont {Arcizet}}, \bibinfo {author} {\bibfnamefont {Mario}\
  \bibnamefont {Ruben}}, \bibinfo {author} {\bibfnamefont {Nedjma}\
  \bibnamefont {Bendiab}}, \ and\ \bibinfo {author} {\bibfnamefont {Vincent}\
  \bibnamefont {Bouchiat}},\ }\bibfield  {title} {\enquote {\bibinfo {title}
  {Time- and space-modulated raman signals in graphene-based optical
  cavities},}\ }\href {http://stacks.iop.org/2040-8986/15/i=11/a=114010}
  {\bibfield  {journal} {\bibinfo  {journal} {Journal of Optics}\ }\textbf
  {\bibinfo {volume} {15}},\ \bibinfo {pages} {114010} (\bibinfo {year}
  {2013})}\BibitemShut {NoStop}%
\bibitem [{\citenamefont {Hencky}(1915)}]{Hencky1915}%
  \BibitemOpen
  \bibfield  {author} {\bibinfo {author} {\bibfnamefont {H.}~\bibnamefont
  {Hencky}},\ }\bibfield  {title} {\enquote {\bibinfo {title} {\"{U}ber den
  Spannungszustand in kreisrunden Platten mit verschwindender
  Biegungssteifigkeit},}\ }\href@noop {} {\bibfield  {journal} {\bibinfo
  {journal} {Z. angew. Math. Phys.}\ }\textbf {\bibinfo {volume} {63}},\
  \bibinfo {pages} {311--317} (\bibinfo {year} {1915})}\BibitemShut {NoStop}%
\bibitem [{\citenamefont {Wan}\ and\ \citenamefont {Mai}(1995)}]{Wan1995}%
  \BibitemOpen
  \bibfield  {author} {\bibinfo {author} {\bibfnamefont {Kai-Tak}\ \bibnamefont
  {Wan}}\ and\ \bibinfo {author} {\bibfnamefont {Yiu-Wing}\ \bibnamefont
  {Mai}},\ }\bibfield  {title} {\enquote {\bibinfo {title} {Fracture mechanics
  of a new blister test with stable crack growth},}\ }\href
  {http://www.sciencedirect.com/science/article/pii/0956715195001088}
  {\bibfield  {journal} {\bibinfo  {journal} {Acta Metallurgica et Materialia}\
  }\textbf {\bibinfo {volume} {43}},\ \bibinfo {pages} {4109--4115} (\bibinfo
  {year} {1995})}\BibitemShut {NoStop}%
\bibitem [{\citenamefont {Thomsen}\ \emph {et~al.}(2002)\citenamefont
  {Thomsen}, \citenamefont {Reich},\ and\ \citenamefont
  {Ordej\'on}}]{Thomsen2002}%
  \BibitemOpen
  \bibfield  {author} {\bibinfo {author} {\bibfnamefont {C.}~\bibnamefont
  {Thomsen}}, \bibinfo {author} {\bibfnamefont {S.}~\bibnamefont {Reich}}, \
  and\ \bibinfo {author} {\bibfnamefont {P.}~\bibnamefont {Ordej\'on}},\
  }\bibfield  {title} {\enquote {\bibinfo {title} {\textit{Ab initio}
  determination of the phonon deformation potentials of graphene},}\ }\href
  {\doibase 10.1103/PhysRevB.65.073403} {\bibfield  {journal} {\bibinfo
  {journal} {Phys. Rev. B}\ }\textbf {\bibinfo {volume} {65}},\ \bibinfo
  {pages} {073403} (\bibinfo {year} {2002})}\BibitemShut {NoStop}%
\bibitem [{\citenamefont {Cheng}\ \emph {et~al.}(2011)\citenamefont {Cheng},
  \citenamefont {Zhu}, \citenamefont {Huang},\ and\ \citenamefont
  {Schwingenschl\"ogl}}]{Cheng2011}%
  \BibitemOpen
  \bibfield  {author} {\bibinfo {author} {\bibfnamefont {Y.~C.}\ \bibnamefont
  {Cheng}}, \bibinfo {author} {\bibfnamefont {Z.~Y.}\ \bibnamefont {Zhu}},
  \bibinfo {author} {\bibfnamefont {G.~S.}\ \bibnamefont {Huang}}, \ and\
  \bibinfo {author} {\bibfnamefont {U.}~\bibnamefont {Schwingenschl\"ogl}},\
  }\bibfield  {title} {\enquote {\bibinfo {title} {Grüneisen parameter of the G
  mode of strained monolayer graphene},}\ }\href {\doibase
  10.1103/PhysRevB.83.115449} {\bibfield  {journal} {\bibinfo  {journal} {Phys.
  Rev. B}\ }\textbf {\bibinfo {volume} {83}},\ \bibinfo {pages} {115449}
  (\bibinfo {year} {2011})}\BibitemShut {NoStop}%
\bibitem [{\citenamefont {Jiang}\ \emph {et~al.}(2009)\citenamefont {Jiang},
  \citenamefont {Wang},\ and\ \citenamefont {Li}}]{Jiang2009}%
  \BibitemOpen
  \bibfield  {author} {\bibinfo {author} {\bibfnamefont {Jin-Wu}\ \bibnamefont
  {Jiang}}, \bibinfo {author} {\bibfnamefont {Jian-Sheng}\ \bibnamefont
  {Wang}}, \ and\ \bibinfo {author} {\bibfnamefont {Baowen}\ \bibnamefont
  {Li}},\ }\bibfield  {title} {\enquote {\bibinfo {title} {Young's modulus of
  graphene: A molecular dynamics study},}\ }\href {\doibase
  10.1103/PhysRevB.80.113405} {\bibfield  {journal} {\bibinfo  {journal} {Phys.
  Rev. B}\ }\textbf {\bibinfo {volume} {80}},\ \bibinfo {pages} {113405}
  (\bibinfo {year} {2009})}\BibitemShut {NoStop}%
\bibitem [{\citenamefont {Tan}\ \emph {et~al.}(2013)\citenamefont {Tan},
  \citenamefont {Wu}, \citenamefont {Zhang}, \citenamefont {Peng},
  \citenamefont {Sun},\ and\ \citenamefont {Zhong}}]{Tan2013}%
  \BibitemOpen
  \bibfield  {author} {\bibinfo {author} {\bibfnamefont {Xinjun}\ \bibnamefont
  {Tan}}, \bibinfo {author} {\bibfnamefont {Jian}\ \bibnamefont {Wu}}, \bibinfo
  {author} {\bibfnamefont {Kaiwang}\ \bibnamefont {Zhang}}, \bibinfo {author}
  {\bibfnamefont {Xiangyang}\ \bibnamefont {Peng}}, \bibinfo {author}
  {\bibfnamefont {Lizhong}\ \bibnamefont {Sun}}, \ and\ \bibinfo {author}
  {\bibfnamefont {Jianxin}\ \bibnamefont {Zhong}},\ }\bibfield  {title}
  {\enquote {\bibinfo {title} {Nanoindentation models and young's modulus of
  monolayer graphene: A molecular dynamics study},}\ }\href {\doibase
  http://dx.doi.org/10.1063/1.4793191} {\bibfield  {journal} {\bibinfo
  {journal} {Applied Physics Letters}\ }\textbf {\bibinfo {volume} {102}},\
  \bibinfo {pages} {071908} (\bibinfo {year} {2013})}\BibitemShut {NoStop}%
\bibitem [{\citenamefont {Ekinci}(2005)}]{Ekinci2005}%
  \BibitemOpen
  \bibfield  {author} {\bibinfo {author} {\bibfnamefont {M.~L.}\ \bibnamefont
  {Ekinci}, \bibfnamefont {K.~L. \&~Roukes}},\ }\bibfield  {title} {\enquote
  {\bibinfo {title} {Nanoelectromechanical systems},}\ }\href
  {http://dx.doi.org/10.1063/1.1927327} {\bibfield  {journal} {\bibinfo
  {journal} {Rev. Sci. Instrum.}\ }\textbf {\bibinfo {volume} {76}},\ \bibinfo
  {pages} {061101} (\bibinfo {year} {2005})}\BibitemShut {NoStop}%
\bibitem [{\citenamefont {Southworth}\ \emph {et~al.}(2009)\citenamefont
  {Southworth}, \citenamefont {Barton}, \citenamefont {Verbridge},
  \citenamefont {Ilic}, \citenamefont {Fefferman}, \citenamefont {Craighead},\
  and\ \citenamefont {Parpia}}]{Southworth2009}%
  \BibitemOpen
  \bibfield  {author} {\bibinfo {author} {\bibfnamefont {D.~R.}\ \bibnamefont
  {Southworth}}, \bibinfo {author} {\bibfnamefont {R.~A.}\ \bibnamefont
  {Barton}}, \bibinfo {author} {\bibfnamefont {S.~S.}\ \bibnamefont
  {Verbridge}}, \bibinfo {author} {\bibfnamefont {B.}~\bibnamefont {Ilic}},
  \bibinfo {author} {\bibfnamefont {A.~D.}\ \bibnamefont {Fefferman}}, \bibinfo
  {author} {\bibfnamefont {H.~G.}\ \bibnamefont {Craighead}}, \ and\ \bibinfo
  {author} {\bibfnamefont {J.~M.}\ \bibnamefont {Parpia}},\ }\bibfield  {title}
  {\enquote {\bibinfo {title} {Stress and silicon nitride: A crack in the
  universal dissipation of glasses},}\ }\href
  {http://link.aps.org/doi/10.1103/PhysRevLett.102.225503} {\bibfield
  {journal} {\bibinfo  {journal} {Phys. Rev. Lett.}\ }\textbf {\bibinfo
  {volume} {102}},\ \bibinfo {pages} {225503--} (\bibinfo {year}
  {2009})}\BibitemShut {NoStop}%
\bibitem [{\citenamefont {Singh}\ \emph {et~al.}(2010)\citenamefont {Singh},
  \citenamefont {Sengupta}, \citenamefont {Solanki}, \citenamefont {Dhall},
  \citenamefont {Allain}, \citenamefont {Dhara}, \citenamefont {Pant},\ and\
  \citenamefont {Deshmukh}}]{Singh2010}%
  \BibitemOpen
  \bibfield  {author} {\bibinfo {author} {\bibfnamefont {Vibhor}\ \bibnamefont
  {Singh}}, \bibinfo {author} {\bibfnamefont {Shamashis}\ \bibnamefont
  {Sengupta}}, \bibinfo {author} {\bibfnamefont {Hari~S}\ \bibnamefont
  {Solanki}}, \bibinfo {author} {\bibfnamefont {Rohan}\ \bibnamefont {Dhall}},
  \bibinfo {author} {\bibfnamefont {Adrien}\ \bibnamefont {Allain}}, \bibinfo
  {author} {\bibfnamefont {Sajal}\ \bibnamefont {Dhara}}, \bibinfo {author}
  {\bibfnamefont {Prita}\ \bibnamefont {Pant}}, \ and\ \bibinfo {author}
  {\bibfnamefont {Mandar~M}\ \bibnamefont {Deshmukh}},\ }\bibfield  {title}
  {\enquote {\bibinfo {title} {Probing thermal expansion of graphene and modal
  dispersion at low-temperature using graphene nanoelectromechanical systems
  resonators},}\ }\href {http://stacks.iop.org/0957-4484/21/i=16/a=165204}
  {\bibfield  {journal} {\bibinfo  {journal} {Nanotechnology}\ }\textbf
  {\bibinfo {volume} {21}},\ \bibinfo {pages} {165204} (\bibinfo {year}
  {2010})}\BibitemShut {NoStop}%
\bibitem [{\citenamefont {Mohanty}\ \emph {et~al.}(2002)\citenamefont
  {Mohanty}, \citenamefont {Harrington}, \citenamefont {Ekinci}, \citenamefont
  {Yang}, \citenamefont {Murphy},\ and\ \citenamefont {Roukes}}]{Mohanty2002}%
  \BibitemOpen
  \bibfield  {author} {\bibinfo {author} {\bibfnamefont {P.}~\bibnamefont
  {Mohanty}}, \bibinfo {author} {\bibfnamefont {D.~A.}\ \bibnamefont
  {Harrington}}, \bibinfo {author} {\bibfnamefont {K.~L.}\ \bibnamefont
  {Ekinci}}, \bibinfo {author} {\bibfnamefont {Y.~T.}\ \bibnamefont {Yang}},
  \bibinfo {author} {\bibfnamefont {M.~J.}\ \bibnamefont {Murphy}}, \ and\
  \bibinfo {author} {\bibfnamefont {M.~L.}\ \bibnamefont {Roukes}},\ }\bibfield
   {title} {\enquote {\bibinfo {title} {Intrinsic dissipation in high-frequency
  micromechanical resonators},}\ }\href
  {http://link.aps.org/doi/10.1103/PhysRevB.66.085416} {\bibfield  {journal}
  {\bibinfo  {journal} {Phys. Rev. B}\ }\textbf {\bibinfo {volume} {66}},\
  \bibinfo {pages} {085416--} (\bibinfo {year} {2002})}\BibitemShut {NoStop}%
\bibitem [{\citenamefont {Bercu}\ \emph {et~al.}(2009)\citenamefont {Bercu},
  \citenamefont {Xu}, \citenamefont {Montès},\ and\ \citenamefont
  {Morfouli}}]{Bercu2009}%
  \BibitemOpen
  \bibfield  {author} {\bibinfo {author} {\bibfnamefont {Bogdan}\ \bibnamefont
  {Bercu}}, \bibinfo {author} {\bibfnamefont {Xin}\ \bibnamefont {Xu}},
  \bibinfo {author} {\bibfnamefont {Laurent}\ \bibnamefont {Montès}}, \ and\
  \bibinfo {author} {\bibfnamefont {Panagiota}\ \bibnamefont {Morfouli}},\
  }\bibfield  {title} {\enquote {\bibinfo {title} {Characterization of
  mechanical stress on nanostructures for NEMS applications by ultra-thin
  membrane and self-suspension techniques},}\ }\href
  {http://www.sciencedirect.com/science/article/pii/S0167931709000112}
  {\bibfield  {journal} {\bibinfo  {journal} {Microelectronic Engineering}\
  }\textbf {\bibinfo {volume} {86}},\ \bibinfo {pages} {1303--1306} (\bibinfo
  {year} {2009})}\BibitemShut {NoStop}%
\bibitem [{\citenamefont {Bauer}\ \emph {et~al.}(2008)\citenamefont {Bauer},
  \citenamefont {Gigler}, \citenamefont {Richter},\ and\ \citenamefont
  {Stark}}]{Bauer2008}%
  \BibitemOpen
  \bibfield  {author} {\bibinfo {author} {\bibfnamefont {M.}~\bibnamefont
  {Bauer}}, \bibinfo {author} {\bibfnamefont {A.M.}\ \bibnamefont {Gigler}},
  \bibinfo {author} {\bibfnamefont {C.}~\bibnamefont {Richter}}, \ and\
  \bibinfo {author} {\bibfnamefont {R.W.}\ \bibnamefont {Stark}},\ }\bibfield
  {title} {\enquote {\bibinfo {title} {Visualizing stress in silicon micro
  cantilevers using scanning confocal raman spectroscopy},}\ }\href
  {http://www.sciencedirect.com/science/article/pii/S0167931708000944}
  {\bibfield  {journal} {\bibinfo  {journal} {Microelectronic Engineering}\
  }\textbf {\bibinfo {volume} {85}},\ \bibinfo {pages} {1443--1446} (\bibinfo
  {year} {2008})}\BibitemShut {NoStop}%
\bibitem [{\citenamefont {Starman~Jr.}\ \emph {et~al.}(2003)\citenamefont
  {Starman~Jr.}, \citenamefont {Lott}, \citenamefont {Amer}, \citenamefont
  {Cowan},\ and\ \citenamefont {Busbee}}]{StarmanJr2003}%
  \BibitemOpen
  \bibfield  {author} {\bibinfo {author} {\bibfnamefont {L.A.}\ \bibnamefont
  {Starman~Jr.}}, \bibinfo {author} {\bibfnamefont {J.A.}\ \bibnamefont
  {Lott}}, \bibinfo {author} {\bibfnamefont {M.S.}\ \bibnamefont {Amer}},
  \bibinfo {author} {\bibfnamefont {W.D.}\ \bibnamefont {Cowan}}, \ and\
  \bibinfo {author} {\bibfnamefont {J.D.}\ \bibnamefont {Busbee}},\ }\bibfield
  {title} {\enquote {\bibinfo {title} {Stress characterization of mems
  microbridges by micro-raman spectroscopy},}\ }\href
  {http://www.sciencedirect.com/science/article/pii/S0924424702004326}
  {\bibfield  {journal} {\bibinfo  {journal} {Sensors and Actuators A:
  Physical}\ }\textbf {\bibinfo {volume} {104}},\ \bibinfo {pages} {107--116}
  (\bibinfo {year} {2003})}\BibitemShut {NoStop}%
\bibitem [{\citenamefont {Xue}\ \emph {et~al.}(2007)\citenamefont {Xue},
  \citenamefont {Zheng}, \citenamefont {Zhang}, \citenamefont {Zhang},\ and\
  \citenamefont {Jian}}]{Xue2007}%
  \BibitemOpen
  \bibfield  {author} {\bibinfo {author} {\bibfnamefont {Chenyang}\
  \bibnamefont {Xue}}, \bibinfo {author} {\bibfnamefont {Lina}\ \bibnamefont
  {Zheng}}, \bibinfo {author} {\bibfnamefont {Wendong}\ \bibnamefont {Zhang}},
  \bibinfo {author} {\bibfnamefont {Binzhen}\ \bibnamefont {Zhang}}, \ and\
  \bibinfo {author} {\bibfnamefont {Aoqun}\ \bibnamefont {Jian}},\ }\bibfield
  {title} {\enquote {\bibinfo {title} {A dynamic stress analyzer for
  microelectromechanical systems (MEMS) based on raman spectroscopy},}\ }\href
  {\doibase 10.1002/jrs.1673} {\bibfield  {journal} {\bibinfo  {journal}
  {Journal of Raman Spectroscopy}\ }\textbf {\bibinfo {volume} {38}},\ \bibinfo
  {pages} {467--471} (\bibinfo {year} {2007})}\BibitemShut {NoStop}%
\bibitem [{\citenamefont {Pomeroy}\ \emph {et~al.}(2008)\citenamefont
  {Pomeroy}, \citenamefont {Gkotsis}, \citenamefont {Zhu}, \citenamefont
  {Leighton}, \citenamefont {Kirby},\ and\ \citenamefont
  {Kuball}}]{Pomeroy2008}%
  \BibitemOpen
  \bibfield  {author} {\bibinfo {author} {\bibfnamefont {J.W.}\ \bibnamefont
  {Pomeroy}}, \bibinfo {author} {\bibfnamefont {P.}~\bibnamefont {Gkotsis}},
  \bibinfo {author} {\bibfnamefont {Meiling}\ \bibnamefont {Zhu}}, \bibinfo
  {author} {\bibfnamefont {G.}~\bibnamefont {Leighton}}, \bibinfo {author}
  {\bibfnamefont {P.}~\bibnamefont {Kirby}}, \ and\ \bibinfo {author}
  {\bibfnamefont {M.}~\bibnamefont {Kuball}},\ }\bibfield  {title} {\enquote
  {\bibinfo {title} {Dynamic operational stress measurement of MEMS using
  time-resolved raman spectroscopy},}\ }\href
  {http://dx.doi.org/10.1109/JMEMS.2008.2004849} {\bibfield  {journal}
  {\bibinfo  {journal} {Microelectromechanical Systems, Journal of}\ }\textbf
  {\bibinfo {volume} {17}},\ \bibinfo {pages} {1315--1321} (\bibinfo {year}
  {2008})}\BibitemShut {NoStop}%
\end{thebibliography}

\begin{thebibliography}{7}%
\makeatletter
\providecommand \@ifxundefined [1]{%
 \@ifx{#1\undefined}
}%
\providecommand \@ifnum [1]{%
 \ifnum #1\expandafter \@firstoftwo
 \else \expandafter \@secondoftwo
 \fi
}%
\providecommand \@ifx [1]{%
 \ifx #1\expandafter \@firstoftwo
 \else \expandafter \@secondoftwo
 \fi
}%
\providecommand \natexlab [1]{#1}%
\providecommand \enquote  [1]{``#1''}%
\providecommand \bibnamefont  [1]{#1}%
\providecommand \bibfnamefont [1]{#1}%
\providecommand \citenamefont [1]{#1}%
\providecommand \href@noop [0]{\@secondoftwo}%
\providecommand \href [0]{\begingroup \@sanitize@url \@href}%
\providecommand \@href[1]{\@@startlink{#1}\@@href}%
\providecommand \@@href[1]{\endgroup#1\@@endlink}%
\providecommand \@sanitize@url [0]{\catcode `\\12\catcode `\$12\catcode
  `\&12\catcode `\#12\catcode `\^12\catcode `\_12\catcode `\%12\relax}%
\providecommand \@@startlink[1]{}%
\providecommand \@@endlink[0]{}%
\providecommand \url  [0]{\begingroup\@sanitize@url \@url }%
\providecommand \@url [1]{\endgroup\@href {#1}{\urlprefix }}%
\providecommand \urlprefix  [0]{URL }%
\providecommand \Eprint [0]{\href }%
\providecommand \doibase [0]{http://dx.doi.org/}%
\providecommand \selectlanguage [0]{\@gobble}%
\providecommand \bibinfo  [0]{\@secondoftwo}%
\providecommand \bibfield  [0]{\@secondoftwo}%
\providecommand \translation [1]{[#1]}%
\providecommand \BibitemOpen [0]{}%
\providecommand \bibitemStop [0]{}%
\providecommand \bibitemNoStop [0]{.\EOS\space}%
\providecommand \EOS [0]{\spacefactor3000\relax}%
\providecommand \BibitemShut  [1]{\csname bibitem#1\endcsname}%
\let\auto@bib@innerbib\@empty
\bibitem [{\citenamefont {Zabel}\ \emph {et~al.}(2011)\citenamefont {Zabel},
  \citenamefont {Nair}, \citenamefont {Ott}, \citenamefont {Georgiou},
  \citenamefont {Geim}, \citenamefont {Novoselov},\ and\ \citenamefont
  {Casiraghi}}]{Zabel2011}%
  \BibitemOpen
  \bibfield  {author} {\bibinfo {author} {\bibfnamefont {Jakob}\ \bibnamefont
  {Zabel}}, \bibinfo {author} {\bibfnamefont {Rahul~R.}\ \bibnamefont {Nair}},
  \bibinfo {author} {\bibfnamefont {Anna}\ \bibnamefont {Ott}}, \bibinfo
  {author} {\bibfnamefont {Thanasis}\ \bibnamefont {Georgiou}}, \bibinfo
  {author} {\bibfnamefont {Andre~K.}\ \bibnamefont {Geim}}, \bibinfo {author}
  {\bibfnamefont {Kostya~S.}\ \bibnamefont {Novoselov}}, \ and\ \bibinfo
  {author} {\bibfnamefont {Cinzia}\ \bibnamefont {Casiraghi}},\ }\bibfield
  {title} {\enquote {\bibinfo {title} {Raman spectroscopy of graphene and
  bilayer under biaxial strain: Bubbles and balloons},}\ }\href {\doibase
  10.1021/nl203359n} {\bibfield  {journal} {\bibinfo  {journal} {Nano Lett.}\
  }\textbf {\bibinfo {volume} {12}},\ \bibinfo {pages} {617--621} (\bibinfo
  {year} {2011})}\BibitemShut {NoStop}%
\bibitem [{\citenamefont {Yoon}\ \emph {et~al.}(2009)\citenamefont {Yoon},
  \citenamefont {Moon}, \citenamefont {Son}, \citenamefont {Choi},
  \citenamefont {Park}, \citenamefont {Cha}, \citenamefont {Kim},\ and\
  \citenamefont {Cheong}}]{Yoon2009}%
  \BibitemOpen
  \bibfield  {author} {\bibinfo {author} {\bibfnamefont {Duhee}\ \bibnamefont
  {Yoon}}, \bibinfo {author} {\bibfnamefont {Hyerim}\ \bibnamefont {Moon}},
  \bibinfo {author} {\bibfnamefont {Young-Woo}\ \bibnamefont {Son}}, \bibinfo
  {author} {\bibfnamefont {Jin~Sik}\ \bibnamefont {Choi}}, \bibinfo {author}
  {\bibfnamefont {Bae~Ho}\ \bibnamefont {Park}}, \bibinfo {author}
  {\bibfnamefont {Young~Hun}\ \bibnamefont {Cha}}, \bibinfo {author}
  {\bibfnamefont {Young~Dong}\ \bibnamefont {Kim}}, \ and\ \bibinfo {author}
  {\bibfnamefont {Hyeonsik}\ \bibnamefont {Cheong}},\ }\bibfield  {title}
  {\enquote {\bibinfo {title} {Interference effect on raman spectrum of
  graphene on SiO$_2$/Si},}\ }\href
  {http://link.aps.org/doi/10.1103/PhysRevB.80.125422} {\bibfield  {journal}
  {\bibinfo  {journal} {Phys. Rev. B}\ }\textbf {\bibinfo {volume} {80}},\
  \bibinfo {pages} {125422--} (\bibinfo {year} {2009})}\BibitemShut {NoStop}%
\bibitem [{\citenamefont {Metten}\ \emph {et~al.}(2013)\citenamefont {Metten},
  \citenamefont {Federspiel}, \citenamefont {Romeo},\ and\ \citenamefont
  {Berciaud}}]{Metten2013}%
  \BibitemOpen
  \bibfield  {author} {\bibinfo {author} {\bibfnamefont {D.}~\bibnamefont
  {Metten}}, \bibinfo {author} {\bibfnamefont {F.}~\bibnamefont {Federspiel}},
  \bibinfo {author} {\bibfnamefont {M.}~\bibnamefont {Romeo}}, \ and\ \bibinfo
  {author} {\bibfnamefont {S.}~\bibnamefont {Berciaud}},\ }\bibfield  {title}
  {\enquote {\bibinfo {title} {Probing built-in strain in freestanding graphene
  monolayers by raman spectroscopy},}\ }\href
  {http://onlinelibrary.wiley.com/doi/10.1002/pssb.201300220/abstract}
  {\bibfield  {journal} {\bibinfo  {journal} {physica status solidi B}\
  }\textbf {\bibinfo {volume} {250}},\ \bibinfo {pages} {2681--2688} (\bibinfo
  {year} {2013})}\BibitemShut {NoStop}%
\bibitem [{\citenamefont {Hencky}(1915)}]{Hencky1915}%
  \BibitemOpen
  \bibfield  {author} {\bibinfo {author} {\bibfnamefont {H.}~\bibnamefont
  {Hencky}},\ }\bibfield  {title} {\enquote {\bibinfo {title} {\"{U}ber den
  spannungszustand in kreisrunden platten mit verschwindender
  biegungssteifigkeit},}\ }\href@noop {} {\bibfield  {journal} {\bibinfo
  {journal} {Z. angew. Math. Phys.}\ }\textbf {\bibinfo {volume} {63}},\
  \bibinfo {pages} {311--317} (\bibinfo {year} {1915})}\BibitemShut {NoStop}%
\bibitem [{\citenamefont {Wan}\ and\ \citenamefont {Mai}(1995)}]{Wan1995}%
  \BibitemOpen
  \bibfield  {author} {\bibinfo {author} {\bibfnamefont {Kai-Tak}\ \bibnamefont
  {Wan}}\ and\ \bibinfo {author} {\bibfnamefont {Yiu-Wing}\ \bibnamefont
  {Mai}},\ }\bibfield  {title} {\enquote {\bibinfo {title} {Fracture mechanics
  of a new blister test with stable crack growth},}\ }\href
  {http://www.sciencedirect.com/science/article/pii/0956715195001088}
  {\bibfield  {journal} {\bibinfo  {journal} {Acta Metallurgica et Materialia}\
  }\textbf {\bibinfo {volume} {43}},\ \bibinfo {pages} {4109--4115} (\bibinfo
  {year} {1995})}\BibitemShut {NoStop}%
\bibitem [{\citenamefont {Bunch}\ \emph {et~al.}(2008)\citenamefont {Bunch},
  \citenamefont {Verbridge}, \citenamefont {Alden}, \citenamefont {van~der
  Zande}, \citenamefont {Parpia}, \citenamefont {Craighead},\ and\
  \citenamefont {McEuen}}]{Bunch2008}%
  \BibitemOpen
  \bibfield  {author} {\bibinfo {author} {\bibfnamefont {J.~Scott}\
  \bibnamefont {Bunch}}, \bibinfo {author} {\bibfnamefont {Scott~S.}\
  \bibnamefont {Verbridge}}, \bibinfo {author} {\bibfnamefont {Jonathan~S.}\
  \bibnamefont {Alden}}, \bibinfo {author} {\bibfnamefont {Arend~M.}\
  \bibnamefont {van~der Zande}}, \bibinfo {author} {\bibfnamefont {Jeevak~M.}\
  \bibnamefont {Parpia}}, \bibinfo {author} {\bibfnamefont {Harold~G.}\
  \bibnamefont {Craighead}}, \ and\ \bibinfo {author} {\bibfnamefont {Paul~L.}\
  \bibnamefont {McEuen}},\ }\bibfield  {title} {\enquote {\bibinfo {title}
  {Impermeable atomic membranes from graphene sheets},}\ }\href {\doibase
  10.1021/nl801457b} {\bibfield  {journal} {\bibinfo  {journal} {Nano Lett.}\
  }\textbf {\bibinfo {volume} {8}},\ \bibinfo {pages} {2458--2462} (\bibinfo
  {year} {2008})}\BibitemShut {NoStop}%
\bibitem [{\citenamefont {Koenig}\ \emph {et~al.}(2011)\citenamefont {Koenig},
  \citenamefont {Boddeti}, \citenamefont {Dunn},\ and\ \citenamefont
  {Bunch}}]{Koenig2011}%
  \BibitemOpen
  \bibfield  {author} {\bibinfo {author} {\bibfnamefont {Steven~P.}\
  \bibnamefont {Koenig}}, \bibinfo {author} {\bibfnamefont {Narasimha~G.}\
  \bibnamefont {Boddeti}}, \bibinfo {author} {\bibfnamefont {Martin~L.}\
  \bibnamefont {Dunn}}, \ and\ \bibinfo {author} {\bibfnamefont {J.~Scott}\
  \bibnamefont {Bunch}},\ }\bibfield  {title} {\enquote {\bibinfo {title}
  {Ultrastrong adhesion of graphene membranes},}\ }\href
  {http://dx.doi.org/10.1038/nnano.2011.123} {\bibfield  {journal} {\bibinfo
  {journal} {Nat Nano}\ }\textbf {\bibinfo {volume} {6}},\ \bibinfo {pages}
  {543--546} (\bibinfo {year} {2011})}\BibitemShut {NoStop}%
\end{thebibliography}
%

\end{document}